\RequirePackage{fix-cm}
\documentclass{svjour3}                     
\smartqed  

\usepackage{algorithmic}
\usepackage{alltt}
\usepackage{amsfonts}
\usepackage{amsmath}
\usepackage{array}
\usepackage{booktabs}
\usepackage{boxhandler}
\usepackage{colortbl}
\usepackage{enumerate}
\usepackage{enumitem}
\usepackage{flushend}
\usepackage{graphicx}
\usepackage{latexsym}
\usepackage{listings}
\usepackage{makecell}
\usepackage{multirow}
\usepackage{proof}
\usepackage{stackengine}
\usepackage{url}
\usepackage{subcaption}
\usepackage{tcolorbox}
\usepackage{textcomp}
\usepackage{url}
\usepackage{xcolor}
\usepackage{xspace}
\usepackage{tabularx}
\usepackage{tikz} 

\newboolean{showcomments}
\setboolean{showcomments}{true}
\ifthenelse{\boolean{showcomments}}
{ \newcommand{\mynote}[2]{
\fbox{\bfseries\sffamily\scriptsize#1}
{\small$\blacktriangleright$\textsf{\textcolor{red}{{\em #2}\bf }}$\blacktriangleleft$}}}
{ \newcommand{\mynote}[2]{}}

\newcommand{\etal}{\textit{et al.}\xspace}
\newcommand{\eg}{\textit{e.g.,}\xspace}
\newcommand{\ie}{\textit{i.e.,}\xspace}

\definecolor{pblue}{rgb}{0.13,0.13,1}
\definecolor{pgreen}{rgb}{0,0.5,0}
\definecolor{pred}{rgb}{0.9,0,0}
\definecolor{pgrey}{rgb}{0.46,0.45,0.48}

\newcommand{\sm}{{\em Software Metrics}\xspace}
\newcommand{\im}{{\em Imports}\xspace}
\newcommand{\fc}{{\em Function Calls}\xspace}
\newcommand{\tm}{{\em Text Mining}\xspace}
\newcommand{\our}{{\em TROVON}\xspace}
\newcommand{\dev}{{\em Devign}\xspace}
\newcommand{\lstm}{{\em LSTM}\xspace}
\newcommand{\lstmrf}{{\em LSTM-RF}\xspace}
\newcommand{\bilstm}{{\em Bi-LSTM}\xspace}
\newcommand{\ourbilstm}{{\em TROVON-BILSTM}\xspace}

\newcommand{\rev}[1]{{#1}}
\newcommand{\minrev}[1]{{#1}}
\newcommand{\norev}[1]{{#1}}
\newcommand{\cleanset}{{\em Clean Training Data Settings}\xspace}
\newcommand{\realset}{{\em Realistic Training Data Settings}\xspace}

\lstdefinestyle{codestyle}{
    commentstyle=\color{pgreen},
    keywordstyle=\color{pblue},
    numberstyle=\scriptsize\color{pgrey},
    stringstyle=\color{pred},
    basicstyle=\ttfamily\small,
    breakatwhitespace=false,         
    breaklines=true,                 
    captionpos=b,                    
    keepspaces=true,                 
    numbers=left,                    
    numbersep=5pt,                  
    showspaces=false,                
    showstringspaces=false,
    showtabs=false,                  
    tabsize=2,
    frame=none
}

\begin{document}
\title{Learning from What We Know
}
\subtitle{How to Perform Vulnerability Prediction using Noisy Historical Data}


\author{Aayush Garg         \and
        Renzo Degiovanni         \and
        Matthieu Jimenez         \and
        Maxime Cordy         \and
        Mike~Papadakis         \and
        Yves Le Traon 
}


\institute{A. Garg \at
              University of Luxembourg \\
              \email{aayush.garg@uni.lu}           
           \and
           R. Degiovanni \at
           University of Luxembourg\\
           \email{renzo.degiovanni@uni.lu}
           \and
           M. Jimenez \at
           University of Luxembourg\\
           \email{matthieu.jimenez@uni.lu}
           \and
           M. Cordy \at
           University of Luxembourg\\
           \email{maxime.cordy@uni.lu}
           \and
           M. Papadakis \at
           University of Luxembourg\\
           \email{michail.papadakis@uni.lu}
           \and
           Y. Le Traon \at
           University of Luxembourg\\
           \email{Yves.LeTraon@uni.lu}
}

\date{}

\maketitle


\begin{abstract}
Vulnerability prediction refers to the problem of identifying system components that are most likely to be vulnerable. 
Typically, this problem is tackled by training binary classifiers on historical data. Unfortunately, recent research has shown that such approaches underperform due to the following two reasons: 
a) the imbalanced nature of the problem, and 
b) the inherently noisy historical data, i.e., most vulnerabilities are discovered much later than they are introduced.
This misleads classifiers as they learn to recognize actual vulnerable components as non-vulnerable. 
To tackle these issues, we propose \our, a technique that learns from known vulnerable components rather than from vulnerable and non-vulnerable components, as typically performed. 
We perform this by contrasting the known vulnerable, and their respective fixed components.
This way, \our manages to learn from the things we know, \ie vulnerabilities, hence reducing the effects of noisy  and unbalanced data. 
We evaluate \our by comparing it with existing techniques on three security-critical open source systems, \ie Linux Kernel, OpenSSL, and Wireshark, with historical vulnerabilities that have been reported in the National Vulnerability Database (NVD).
Our evaluation demonstrates that the prediction capability of \our significantly outperforms existing vulnerability prediction techniques such as \sm, \im, \fc, \tm, \rev{\dev, \lstm, and \lstmrf} with \rev{an improvement of 40.84\% in \emph{Matthews Correlation Coefficient}~(MCC) score under \cleanset, and an improvement of 35.52\% under \realset.} 

\end{abstract}


\section{Introduction}
\label{sec:introduction}

A vulnerability is a hole or a weakness in the application, which can be a design flaw or an implementation bug, that allows an attacker to cause harm to the stakeholders, \ie the application owner, application users, and other entities that rely on the application~\cite{Vulnerabilities}.
While vulnerabilities can be thought of as specific types of software defects (or bugs), 
there are subtle and significant differences that make their identification considerably more complex and challenging than the problem of finding bugs \cite{TangZYLZX15,PotterMcGraw2004}. 

Vulnerabilities  are fewer in comparison to defects, limiting  the information one can learn from. Also, their identification requires an attacker's mindset \cite{Morrison+2015}, which developers or code reviewers may not possess.
Lastly, the continuous growth of codebases makes it difficult to investigate them entirely and track all code changes. 
For example, the Linux kernel, one of the projects with the highest number of publicly reported vulnerabilities, reached \rev{27.80} million LoC (Lines of Codes) at the beginning of 2020~\cite{linux-news}.
 
Vulnerability prediction approaches were proposed to tackle these challenges by prioritizing the efforts that developers and code reviewers have to put on when testing or reviewing code to find vulnerabilities. These methods \rev{take advantage of the large amounts of historical data available based on which they learn} a set of features and/or code properties that associate with vulnerabilities. For instance, the presence of vulnerabilities has been linked to high code churn~\cite{ShinMWO2011}, 
to the use of specific library imports and function calls~\cite{Neuhaus+2007}, and the frequency of suspicious code tokens ~\cite{TangZYLZX15}. 
Unfortunately, building models around such features is challenging due to the small number of available vulnerable code instances, which limit the learning ability of the predictors~\cite{Zimmermann+2009}. 

Furthermore, \emph{Jimenez}~\etal\cite{Jimenez+2019} demonstrated that vulnerability prediction approaches have been built under a \emph{``clean''} training data assumption, \ie all the component's labeling information (vulnerable~/~non-vulnerable) is always available irrespective of time.
Their study showed that under these settings the approaches do not account for the gradual revelation of vulnerabilities over time.
This results in prediction models training on even those vulnerabilities that have not been uncovered yet, 
\eg all vulnerabilities known from time \emph{t} onwards are available at all times, even before time \emph{t}.

Jimenez~\etal advocated \rev{\realset} where the vulnerability labels used for training the prediction models are more realistically available at training time.
For example, in such settings, at a given time \emph{t}, only the vulnerabilities known till time \emph{t} should be available for training.
All vulnerabilities known from time \emph{t} onwards should \emph{not} be available for training beforehand.
Their study demonstrated that \realset results in unavoidable noise in the training data because every component with no reported vulnerability till training time is considered non-vulnerable during training, which makes existing approaches perform poorly.
This establishes a need for robust vulnerability prediction techniques.

We advance in this direction by developing \our\footnote{\textbf{\our} is an abbreviation for ``\textbf{Tr}aining \textbf{o}n \textbf{v}ulnerabilities \textbf{on}ly'', which is the core focus of our study.}-- a method that learns from validated data, i.e., we train only on components known to be vulnerable and leave aside the (supposedly) non-vulnerable ones. 
This way, we do not make any assumptions on non-vulnerable components and bypass the key problem faced by previous works. To do so,  we rely on a simple yet powerful language-agnostic machine translation technique \cite{Britz:2017} which we train on pairs of vulnerable and fixed code fragments, available at projects' release time. \rev{In particular, we contrast the code fragments pairs (pairs of vulnerable and fixed fragments) that were modified when fixing a vulnerability, with fragment pairs from other functions of the same components (fragments less likely to be vulnerable) in order to learn to distinguish likely vulnerable from non-vulnerable code.}

\our focuses on vulnerability fixes, i.e., code transformations that turn vulnerable code into a non-vulnerable one, to train the machine translation model that aims at capturing silent features related to the differences between vulnerable and fixed components.  
Therefore, predictions are guided by actual points of interest, (\ie diff points) in the vulnerable code where the transformations should happen. 
This means that \our learns to identify code characteristics that are similar to those (vulnerable) seen during training.  

We empirically assess the effectiveness of \our on available releases of three security-critical open source systems, \ie Linux Kernel, Wireshark, and OpenSSL.
Our evaluation demonstrates that \our significantly outperforms existing vulnerability prediction \rev{approaches} under both \rev{\cleanset and \realset}.

\rev{In particular, our results show that when we train all the approaches (including \our) with clean training data, \our outperforms the existing approaches by 83.96\% in Precision, 155.33\% in Recall, 132.95\% in F-measure, and 80.39\% in Matthews Correlation Coefficient (MCC)}.
In addition to these metrics, we also evaluate \our on predicting unseen vulnerable components specifically. 
This is a new metric that we introduce in this paper to help evaluate the extent to which vulnerability prediction generalizes, i.e., \rev{ability to predict unseen components  (components not used for training) as being vulnerable or not}. 
The percentages of \rev{unseen} vulnerable components predicted by TROVON, on average, are \rev{40.05\%}, \rev{64.34\%}, and \rev{42.28\%} higher than the ones obtained by existing techniques in Linux Kernel, Wireshark, and OpenSSL releases, reflecting \our's better generalization capability.
\rev{Under \realset, on average, TROVON achieved 0.39 MCC, (\ie 3.63 times higher than the baselines), 0.69 F-measure, (\ie 11.82 times higher), 0.86 Precision, (\ie 2.66 times higher), and 0.58 Recall, (\ie 15.25 times higher than the baselines).}

In summary, we make the following contributions:
\begin{enumerate}
    \item We present \our, a novel vulnerability prediction method via machine translation. 
    \item We demonstrate that \our significantly outperforms existing methods through a large empirical study.
    \item We corroborate that \our remains robust when trained in \realset that includes unavoidable noise, where \rev{almost all previous methods that we compared with,} fail~\cite{Jimenez+2019}.
\end{enumerate}

\section{Background}
\label{sec:background}

\subsection{Vulnerabilities}
\label{subsec:background_vulnerabilities}
Common Vulnerability Exposures (CVE)~\cite{CVE:Terminology} defines a security vulnerability as ``\emph{a mistake in software that can be directly used by a hacker to gain access to a system or network}''.
The inadvertence of a developer or insufficient knowledge of defensive programming usually causes these mistakes. 
Still, vulnerabilities are of critical importance for software vendors, who often offer bounties to find them and prioritize their resolution over other less harmful bugs, hence reducing a potential business impact.

Vulnerabilities are usually reported in publicly available databases to promote their disclosure and fix. \minrev{One such example is National Vulnerability Database, aka NVD~\cite{NVD}. NVD is the U.S. government repository of standards based vulnerability management data. All vulnerabilities in the NVD have been assigned a CVE (Common Vulnerabilities and Exposures) identifier. The Common Vulnerabilities and Exposures (CVE) Program’s primary purpose is to uniquely identify vulnerabilities and to associate specific versions of codebases (e.g., software and shared libraries) to those vulnerabilities. The use of CVEs ensures that two or more parties can confidently refer to a CVE identifier (ID) when discussing or sharing information about a unique vulnerability. For every vulnerability, along with the Git commit IDs of the code related to vulnerability-fix commit, NVD also provides related information, i.e., CVE number, vulnerability description, CWE number (if applicable), time of creation, and the list of the impacted releases in the form of reports.}

\subsection{Vulnerability Prediction Modeling}
\subsubsection{Prediction modeling} Prediction modeling aims at learning statistical properties of interest based on historical data. While the resulting models are usually suitable only for the project/application on which they have been trained, the learning process is generic and applies to a specific set of features that associate with the property to predict. In the context of vulnerabilities, a prediction model can be used to classify software components as likely or unlikely vulnerable. 
This information can be used to support the code review process. 
The task is similar to defect prediction, yet due to the sparsity of available examples, it is harder to predict vulnerabilities than defects \cite{shin_can_2013,TheisenW20}.

\subsubsection{Intra vs Inter Predictions:}
Prediction modeling is usually performed in both intra- and cross-project fashion, i.e., training on data of the same or of other projects. 
However, vulnerabilities are project-specific, i.e., they are tied to the project context, used libraries, and development process, and thus, inter-project predictions do not work. 
\rev{Scandariato \etal~\cite{6860243} found that the models for 11 apps out of 20 were too specific for cross-project prediction, and that the link was more pairwise rather than generic. The results of cross-project vulnerability prediction in the study of Sara~\etal~\cite{DBLP:conf/sac/MoshtariS16} show high recall but comparatively low F2 using coupling and IVH.}
Therefore, research in this area is focussed on intra-project.

\subsection{Granularity Level}
Prediction models can target various levels of granularity, such as line, function, component, etc. 
However, the key target should be actionable for the developers and code reviewers that are envisioned to use the technique. 
Given this, a commonly accepted tradeoff is the component (file) level granularity as it has been vetted by Microsoft developers in a study of Morrison \etal \cite{Morrison+2015}, and is used by most existing approaches.
Thus, we consider a code-file as our component, \ie file-level granularity, as it is actionable for industrial use \cite{Morrison+2015}, and provides a baseline for comparing our results with those reported in the relevant literature that we elaborate in Section~\ref{subsec:benchmarks}.

\subsection{\cleanset}
\emph{Jimenez}~\etal\cite{Jimenez+2019} demonstrated that the existing vulnerability prediction approaches have been built under a \emph{``clean''} training data assumption, \ie all the component's labeling information (vulnerable~/~non-vulnerable) is always available irrespective of time, which is unrealistic. Jimenez \etal showed that under these settings, aka \cleanset, prediction approaches fail to account for the gradual revelation of vulnerabilities over time. This results in biased prediction models, i.e., models trained on vulnerabilities that have not been been discovered at the release time, \eg all vulnerabilities known from time \emph{t} onwards are available at all times, even before time \emph{t}.

\subsection{\realset}
In contrast to \cleanset, where the component's labeling information (vulnerable~/~non-vulnerable) is always available irrespective of time, \realset necessitate vulnerability labels to be used for training the prediction models  to be those that are available at training time. For instance in \realset, at a given time \emph{t}, only the vulnerabilities known at time \emph{t} should be available for training. All vulnerabilities known after time \emph{t} should \emph{not} be available for training beforehand. Jimenez \etal study demonstrated that \realset introduce noise in the training data, because every component with no reported vulnerability till the training time is considered as non-vulnerable during training, that makes existing approaches perform poorly.

Irrespective of the poor performance of existing approaches, \realset represents a realist case study, the vulnerabilities are discovered and fixed long after the release date of the projects. In our release-based experiments, (\ie one release for training the model and next release for testing the trained model), only those components are considered as vulnerable in the training set whose vulnerabilities have been discovered and fixed before the next release date of the system.
\subsection{Seen Vulnerable Components}
Vulnerabilities can remain in the code and get propagated throughout different releases (one release after another) of a system, without getting fixed. Due to this, in a release-based experiment, (\ie one release for training the model and next release for testing the trained model), vulnerable components that are present in the training set and \emph{``seen"} by the prediction model during training can also appear in the testing set. Throughout our paper, we refer to such components as \emph{Seen} vulnerable components.

\subsection{Unseen Vulnerable Components}
From one release of a system to the next one, many files/components are modified either to introduce a new functionality or to modify an existing one. In case of Linux Kernel, Wireshark, and OpenSSL projects, we analyzed that 29.95\%, 72.53\%, and 73.58\% of the files, on average, are changed between the releases. A component that was non-vulnerable in the previous release can be become vulnerable in this release, because of such a modification by a developer. Due to this, in a release-based experiment, any component in a testing set which is vulnerable and is not available in the training set, represents a novel vulnerability. Since, this component is \emph{``unseen"} and has not been trained on by the model, we refer to it as \emph{Unseen} vulnerable component.

\subsection{Machine Translation}
\label{subsec:machine-translation}
We perform vulnerability prediction using Machine Translation. 
Machine Translation can be considered as a transformation function $\mathit{transform(X) = Y}$, where the input $\mathit{X = \{x_1, x_2, \ldots, x_n\}}$ is a set of \emph{entities}
that represents a component to be transformed to produce the output $\mathit{Y = \{y_1, y_2, \ldots, y_n\}}$, which is a set of entities that represent a desired component.

In the training phase, the transformation function learns on the example pairs $\mathit{(X,Y)}$ available in the training dataset.
In our context, $X$ contains vulnerable entities, representing a vulnerable component, and $Y$ contains fixed entities, representing the corresponding fixed component.
The transformation function can be trained \emph{not} to transform, \ie to \rev{reproduce} the same output as the input in cases where $X$ is the desired entity-set.
This is achieved by training the function on the example pairs $(X,X)$, i.e. $\mathit{transform(X) = X}$.
In the case of vulnerability prediction modeling, this learned transformation will be used as our prediction model.

\subsection{RNN Encoder-Decoder architecture}
\label{subsec:encoder-decoder-architecture}
\rev{The encoder-decoder architecture for recurrent neural networks is the standard neural machine translation method that rivals and in some cases outperforms classical statistical machine translation methods~\cite{rnnencoderdecoder}. We use the RNN Encoder-Decoder that is established and is used by many recent studies~\cite{garg2022cerebro,sutskever2014sequence,Tufano_2019}.}
The RNN Encoder-Decoder machine translation is composed of two major components: \emph{RNN Encoder} to encode a sequence of terms $x$ into a vector representation, and \emph{RNN Decoder} to decode the representation into another sequence of terms $y$.
The model learns a conditional distribution over an output sequence conditioned on another input sequence of terms: $P(y_1; \ldots; y_m|x_1; \ldots; x_n)$, where $n$ and $m$ may differ. 
For example, given an input sequence $x$ = $Sequence_{in}$ = $(x_1; \ldots; x_n)$ and a target sequence $y$ = $Sequence_{out}$ =  $(y_1; \ldots; y_m)$, the model is trained to learn the conditional distribution: $P(Sequence_{out}|Sequence_{in}) = P(y_1; \ldots; y_m|x_1; \ldots; x_n)$, where $x_i$ and $y_j$ are separated tokens.
A bi-directional RNN Encoder~\cite{Britz:2017}, formed by a backward RNN and a forward RNN, is considered the most efficient to create representations as it takes into account both past and future inputs while reading a sequence~\cite{bahdanau2014neural}.

\section{Approach}
\label{sec:approach}

The key idea of \our is to train a machine translator (viz. an encoder-decoder sequence to sequence model) to identify vulnerable code, 
by feeding it with vulnerable code fragments and their corresponding fixes.  
Machine translators can automatically recognize: 
(i) features of the language (to be translated) and (ii) required translation (to the desired language). 
In our case, it is used to automatically identify vulnerability features with minimum overhead.

It should be noted that we do not aim at fixing vulnerable code, but rather at identifying likely vulnerable code instances. 
The point here is that we use the translator to indicate the presence of vulnerabilities without considering the fixes produced by the model. 
In other words, we leverage the ability of the translators to learn the vulnerabilities' context and not their instance and location.
We assert that since vulnerable code instances are scarce, information gained from historical data is inevitably partial and incomplete. 
Therefore, it can be used to indicate the presence of vulnerabilities but not their instance context. 

The translator is trained on \emph{input - desired output} pairs,\ie on \emph{vulnerable - fixed} code fragments.
For prediction, one can input an unseen code into the trained translator to check whether it is likely to be vulnerable. 
If the translator changes the code then it can be concluded that the code is likely to be vulnerable. 
To avoid many false positives (the translator changing every input code fragment), we also train it to leave non-vulnerable code fragments \emph{unchanged}. 
To this end, we also feed the translator with input-output pairs where each of which is a non-vulnerable code fragment (input = output). 
It must  be noted that we train only on the components (files) that were fixed, leaving aside the unchanged ones. 
This way we aim at reducing the noise from the training data,~\ie by focusing on what we are certain of; the information provided by the vulnerability fixes. 

\begin{figure*}[t]
\begin{center}
\includegraphics[width=\textwidth]{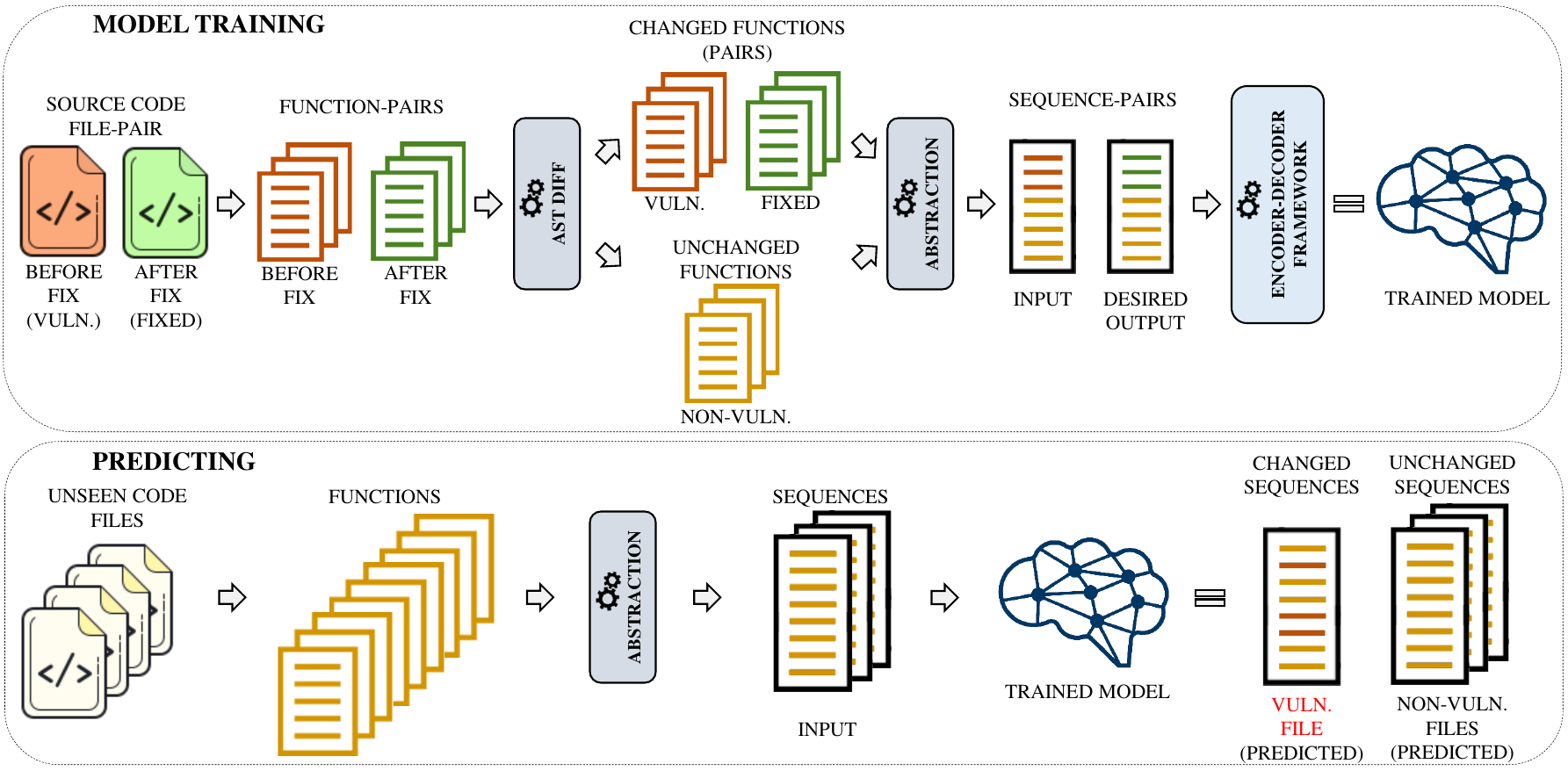}
\caption{Implementation: Sequences generated from the source-code are used to train the model to generate desired output sequences. The trained model is provided with sequences generated from an unseen source code. The component prediction is based on the generated output sequences.}
\label{fig:implementation}
\vspace{-0.5em}
\end{center}
\end{figure*}

Figure~\ref{fig:implementation} shows an overview of the implementation.
Starting from vulnerable code components and their fixes, it involves the following activities: 
1) decomposing the components into code fragments; 
2) identifying which code-fragments are responsible for the vulnerability; 
3) producing abstracted code-fragments by removing irrelevant information (e.g. user-defined names, comments); 
4) configuring and training the machine translator.
5) producing abstracted code-fragments of an unseen code component and using the trained machine translator to predict whether it is likely to be vulnerable.

\subsection{Decomposing Components into Code Fragments}

We target our predictions at the component, (\ie file) level due to: a) the empirical evidence provided by Morrison \etal \cite{Morrison+2015} and b) to account for the context of code (vulnerability-fixes) that can be fixed at multiple locations throughout the component.
A code-fix can be an addition, removal, and/or modification of code. 
Since functions are the basic building blocks of a program, we use them to establish function-level mappings between the vulnerable components and their fixed counterparts (based on the function headers).

Thus, we extract all the functions from both, a vulnerable component and its fixed counterpart, and pair each before-fix function with the corresponding after-fix function. 
The functions that cannot be paired, i.e., having no counterpart, are discarded. 
This can happen due to the creation and/or deletion of a function to fix a vulnerability, \eg a function added during the fix which was not present before or vice-versa.

\subsection{Categorizing Functions as Vulnerable or Non-Vulnerable}

As typically performed in this line of work, we consider as vulnerable, any function that was modified to fix the vulnerability. 
The remaining are considered as non-vulnerable (not vulnerable to the specific vulnerability).
When comparing a before-fix copy to its after-fix counterpart, we ignore irrelevant syntactical changes, \eg additional blank spaces and new lines. 
If there remain syntactical differences between the two copies, we label the before-fix as vulnerable.

\subsection{Abstracting Irrelevant Information}

A major challenge in dealing with raw source code is the huge vocabulary created by the abundance of identifiers and literals used in the code.
Vocabulary, on such a large scale, hinders the learning of relevant code patterns~\cite{Tufano_2019}.
Thus, to reduce the vocabulary size, we transform the source code into an abstract representation by replacing user-defined entities with re-usable IDs.

\begin{figure*}[t]
  \begin{subfigure}{0.49\textwidth}
    \begin{lstlisting}[language=C]
void dev_load (struct net *netw, 
    const char *name ) {
  struct net_device *dev; 
  rcu_read_lock(); 
  dev = dev_get_by_name_rcu (
    netw , name); 
  rcu_read_unlock(); 
  if ( !dev && capable (
      CAP_NET_ADMIN ) )
   request_module("%s", name); 
}
\end{lstlisting}
    \caption{Actual Function}
    \label{fig:real-function}
  \end{subfigure}
  \hfill
  \begin{subfigure}{0.46\textwidth}
    \begin{lstlisting}[language=C]
void F_1 (struct T_1 *V_1, 
    const char *V_2 ) {
  struct T_2 *V_3; 
  F_2(); 
  V_3 = F_3 (
    V_1, V_2);
  F_4(); 
  if ( !V_3 && F_5 (
      V_4 ) ) 
   F_6(L_1, V_2); 
}
\end{lstlisting}
    \caption{Abstracted Function}
    \label{fig:abstracted-function}
  \end{subfigure}
  \caption{Abstraction: Actual Functions (left) are abstracted by replacing user-defined Function names, Type names, Variable names, and String Literals to F\_num, T\_num, V\_num, and L\_num, respectively to achieve Abstracted Functions (right).}
  \label{fig:abstraction}
\end{figure*}

Figure \ref{fig:abstraction} shows a code snippet of a real function (Figure~\ref{fig:real-function}) converted into its abstract representation (Figure~\ref{fig:abstracted-function}).
The purpose of this abstraction is to replace any reference to user-defined entities (function name, type name, variable name, and string literal) with IDs that can be reused across functions (thereby reducing vocabulary size). 
Thus, we replace identifiers and string literals with unique IDs. 
Additionally, comments and annotations are removed.

New IDs follow the regular expression \texttt{(F|T|V|L)\_(num)$^+$}, where \texttt{num} stands for numbers $0, 1, 2, \ldots$ assigned in a sequential and positional fashion based on the occurrence of that entity.
All the entities - user-defined \emph{Function} names, \emph{Type} names, \emph{Variable} names, and \emph{String Literals} are replaced with \texttt{F\_num}, \texttt{T\_num}, \texttt{V\_num}, and \texttt{L\_num}, respectively.
Thus, the first function name receives the ID \texttt{F\_1}, the second receives the ID \texttt{F\_2}, and so on.
If any of these entities appear multiple times in a function, it is replaced with the same ID.

Each function (pair) is abstracted in isolation to yield an abstracted function code, \ie same IDs can be reused across functions without impacting \our. 
ID references are not preserved across functions, e.g., \texttt{V\_1} may refer to two different variable names from one function to another.
This is the key to reduce the vocabulary size, \eg the name of the first function called in any pair is replaced with the ID \texttt{F\_1}, regardless of its original name.

In the case of vulnerable functions, the before-fix copy is abstracted first and then the after-fix copy. IDs are shared between the two copies (before-fix and after-fix) of the functions and new IDs are generated only when new (\emph{Function}, \emph{Type}, \emph{Variable}) names and \emph{String Literals} are found. 

The abstracted code is rearranged in a single sentence to represent a sequence of space-separated entities, which is the representation supported by the machine translator. 
Sequences generated from vulnerable (before-fix), fixed (after-fix), and unchanged functions are named vulnerable, fixed, and unchanged sequences, respectively. 
In these settings, fixed and unchanged sequences represent non-vulnerable cases.
To limit the computation cost involved in training the translator, large sequences are split into multiple sequences of no more than \textit{50} tokens each.

\subsection{Building the Machine Translator}
\label{subsec:Configuring Encoder-Decoder architecture}

To build our machine translator, we train an encoder-decoder model that can transform an input sequence to the desired sequence (output of the model). 

A representation of a sequence is similar to a sentence in a natural language that consists of words separated by spaces and ends with a full stop. 
Instead of words and full stop character, a sequence has tokens and a newline character. 
Thus, we train the encoder-decoder by feeding it with pairs of sequences. 
More precisely, we use two types of pairs: 
(i) vulnerable sequences with their corresponding fixed sequences, and
(ii) non-vulnerable sequences paired with themselves.
Non-vulnerable sequence-pairing is essential to allow the learner to identify what should not be changed.
Thereby, avoiding to raise many false positives (incorrectly predicting non-vulnerable sequences as vulnerable) while learning only from ``clean" data.

\subsection{Predicting Vulnerable Components}
\label{subsec:predicting}

To predict whether an unseen component, (\ie file) is potentially vulnerable, we decompose it into sequences following the process depicted in Figure \ref{fig:implementation}.
Then, we feed the resulting sequences into the machine translator which produces output sequences. 
If one (or more) of the output sequences returned by the model is different from the original one, (\ie input sequences), we consider the component as likely to be vulnerable.
Otherwise, we consider component as likely non-vulnerable, \ie in case of no change in any of the output sequences in comparison to the input sequences, the component is considered as likely non-vulnerable.

\section{Experimental Evaluation}
\label{sec:validation}

\subsection{Research Questions}
\our aims to support code reviews by predicting vulnerable components in new releases, based on the information learned from previous (historical) data, i.e., the previous project release. Therefore, our first research question regards the prediction ability of \our. \rev{We measure the prediction ability of \our to correctly predict vulnerable and non-vulnerable components. We do so with the help of classification assessment metrics, i.e., Precision, Recall, F-1, and MCC}. We evaluate this by training on all available vulnerabilities of one release and testing on the next release, for all available \rev{release} pairs. Thus, we ask:

\begin{itemize}
  \item[\textbf{RQ1}] \rev{What is the prediction performance of \our in a release-based scenario~?}
\end{itemize}

After assessing the prediction \rev{ability} of \our, we turn our attention to existing techniques. Hence, we investigate:

\begin{itemize}
  \item[\textbf{RQ2}] \rev{What is the prediction performance of} \our in comparison to existing techniques?
\end{itemize}

In \our, we train a model on the vulnerabilities of a release and test the trained model on the components of the next release. \rev{Since we perform a release-based evaluation, vulnerabilities spanning across multiple releases could be either seen by the trained model (used during training) or not (newly appearing component).} Thus, we may have the knowledge in advance that a component is vulnerable in a given release irrespective of the vulnerability detection date. As these vulnerable components may remain unfixed and reappear in the next release, it is essential to assess the learning potential of our models by evaluating \rev{how proficient are the studied models in classifying correctly components that were ``seen'' during training, in a sense checking how well the model remembers, and in classifying new components, i.e., components that were ``unseen'' during training, in a sense checking how well a model can actually perform on new instances.} Hence, we aim at controlling for seen and unseen vulnerable components and ask:

\begin{itemize}
  \item[\textbf{RQ3}] What is the prediction performance of the studied techniques in predicting seen and unseen vulnerable components?
\end{itemize}

Until now, we consider that in every release all known vulnerable components are labelled as such, \ie following the \emph{clean} training data settings.
This analysis provides indications on what the potential prediction ability of the approaches is when the available data are clean, \ie all the component’s labeling information (vulnerable / non-vulnerable) is always available irrespective of time. 
Unfortunately, in practice, such information is unavailable and inflates the actual performance of the prediction models. The actual performance in \rev{\realset} is much lower due to real-world labeling issues~\cite{Jimenez+2019}, \ie vulnerabilities are frequently reported at a much later time than they are actually introduced.
This has adverse effects as they cause the classifiers to treat vulnerable components as non-vulnerable. 
Hence, it is imperative to study performance under \rev{\realset}, where a prediction model is trained only on those vulnerabilities that were detected till the release date of a version for which the vulnerability prediction is performed. 
For this reason, we also evaluate the approaches under \rev{\realset}. Hence, we ask:

\begin{itemize}
  \item[\textbf{RQ4}] How effective (in predicting vulnerable components) is \our in comparison to existing techniques under \rev{\realset}?
\end{itemize}

\subsection{Data}
\label{subsec:datasets}
For our study, we need projects with many releases and vulnerabilities. 
We consider three large security-intensive open-source systems that were used by previous research~\cite{Jimenez+2019} -- the Linux Kernel, the OpenSSL library, and the Wireshark tool.
These systems are widely used, mature, and have a long history of releases and vulnerability reports.

\emph{Linux Kernel}~\cite{LinuxKernel} is an operating system, integrated into billions of systems and devices, such as Android.
Linux is one of the largest open-source code-bases and has a long history (since 1991), recorded in its repository.
It is relevant for our evaluation since it has many security aspects and is among the projects with a higher number of reported vulnerabilities in NVD.
\emph{OpenSSL}~\cite{OpenSSL} is a library implementing the SSL and TLS protocols, commonly used in communications.
It is of critical importance as highlighted by the \emph{Heartbleed} vulnerability, which made half of a million web servers vulnerable to attacks~\cite{heartbleed}.
\emph{Wireshark}~\cite{WireShark} is a network packet analyzer mainly used for troubleshooting and debugging.
The project is open source and is relevant for the study because it is integrated with most operating systems.

We use \emph{VulData7}~\cite{JimenezPT18} which is a publicly available\footnote{\url{https://github.com/electricalwind/data7}} tool to gather the vulnerabilities, \ie the vulnerable and the corresponding fixed components of the aforementioned systems.
\minrev{
As we mention in section~\ref{subsec:background_vulnerabilities}, for every vulnerability, NVD provides a Git commit IDs of the code related to vulnerability-fix commit. By using these NVD provided Git commitIDs, \emph{VulData7} extracts the code of vulnerabilities, (i.e., vulnerable code and its patch) and creates a vulnerability dataset.
}

To gather the code-base of these systems, we use \emph{FrameVPM}~\cite{Jimenez+2019} which is also a publicly available tool\footnote{\url{https://github.com/electricalwind/framevpm}}.
\emph{FrameVPM} is a framework built to evaluate and compare vulnerability prediction models.
We also used \emph{FrameVPM} to perform a prediction comparison with existing techniques. 
Section~\ref{subsec:benchmarks} elaborates on the re-implementation of existing techniques that we compare with.
Table \ref{tab:datasetstatistics} provides the details of our dataset.
The dataset composed of the vulnerabilities reported in National Vulnerability Database (NVD)~\cite{NVD}, and the codebase gathered for the 36 releases of Linux Kernel project~\cite{LinuxKernel}, 10 releases of Openssl project~\cite{OpenSSL}, and 10 releases of Wireshark project~\cite{WireShark} is publicly available\footnote{\url{https://github.com/garghub/TROVON}}\rev{, along with our source-code and our re-implemented source-code of the baselines that we compared \our with}.

\begin{table}[t]
  \begin{center}
    \caption{The table records the total number of \rev{releases}, average number of components,  average number of vulnerable components, and the ratio of vulnerable components for the systems we study.}
    \vspace{-1.0em}
    \label{tab:datasetstatistics}
    \begin{tabular}{|l|r|r|r|r|}
    \hline
      \rule{0pt}{3ex}\textbf{System} & \textbf{\#\rev{Releases}} & \textbf{\#Avg.Comp} & \textbf{\#Avg.Vuln.Comp} & \textbf{\%Vuln.} \\
      \hline
      \rule{0pt}{3ex}Linux Kernel & 36 & 16456 & 456 & 3\% \\
      \hline
      \rule{0pt}{3ex}Wireshark & 10 & 2012 & 134 & 7\% \\
      \hline
      \rule{0pt}{3ex}OpenSSL & 10 & 664 & 59 & 9\% \\
      \hline
    \end{tabular}
  \end{center}
\vspace{-0.5em}
\end{table}

\subsection{Implementation and Model Configuration}

During the abstraction phase, we rely on the \emph{srcML} tool~\cite{7816536} to convert source code into an XML format including tags to identify
literals, keywords, identifiers, and comments. 
This helps in separating user-defined identifiers and string literals (the largest part of the vocabulary) from language keywords (a limited set). 
Then, ID replacement is performed by a dedicated tool that we implemented. 
To check whether before and after-fix copies are different, we input the XML produced by \emph{srcML} into the \emph{Gumtree Spoon AST Diff}~\cite{falleri:hal-01054552} tool.
The purpose of using \emph{Gumtree Spoon AST Diff} is to achive a fine-grained diff which can ignore irrelevant changes such as whitespaces and/or new line characters.
It should be noted that \our is not bound to the above-mentioned third-party tools. 
As an alternative, one can use any utility that identifies user-defined entities and performs a diff.

Our encoder-decoder model is built on top of \emph{tf-seq2seq}~\cite{tensorflow2015-whitepaper}, a general-purpose encoder-decoder framework.
To configure it, we learn from previous works that apply machine translation to solve software engineering tasks other than vulnerability prediction, \eg~\cite{Tufano_2019,DBLP:journals/tosem/TufanoWBPWP19,garg2022cerebro}. 
Thus, we rely on a bidirectional encoder as it generally outperforms a unidirectional encoder~\cite{bahdanau2014neural}. 
We use a Long Short-Term Memory (LSTM) network~\cite{HochreiterHochreiter1997} to act as the Recurrent Neural Network (RNN) cell, which was shown to perform better than other common alternatives like simple RNNs or gated recurrent units, in other software engineering prediction tasks~\cite{lstmcomparison2019,lstmcomparisonwebsite}. 
Bucketing and padding are used to deal with the variable length of sequences. 
To strike a balance between performance and training time, we utilize AttentionLayerBahdanau as our attention class, configured with 2 layered AttentionDecoder and 1 layered BidirectionalRNNEncoder, both with 256 units.

To determine an appropriate number of training steps, we conducted a preliminary study involving a validation set (independent of both the training set and the test set that we use in our experimental evaluation) and trained the model by iterations of 5,000 steps. 
At the end of each iteration, we check whether the prediction accuracy on the validation set improved.
If it improved, then we pursued the training for another iteration, otherwise, stopped. 
We found out that the model stopped improving at 50,000 steps, which we thus set as a threshold. 
This order of magnitude is in line with previous research applying machine translation to solve software engineering prediction tasks, e.g., \cite{garg2022cerebro,Tufano_2019}.

\subsection{Experimental Settings}
\label{subsec:experimentalsettings}

Our experimental evaluation is designed to evaluate techniques under \rev{\cleanset and \realset}.
We train a model on each release and test the trained model on the \rev{following} release, (\rev{\ie next} release) simulating a typical release-based vulnerability prediction evaluation scenario~\cite{Jimenez+2019}.

\textit{\rev{\cleanset} - Used in RQs 1, 2 \& 3: }
In these settings, a prediction model is trained using all the vulnerabilities (\rev{vulnerable, \ie} before-fix sequences transformation to \rev{non-vulnerable, \ie} after-fix sequences) of a release of a system (Linux Kernel / OpenSSL / Wireshark).
The trained models are evaluated based on their predictions in the following release of the same system (e.g., trained on vulnerable components in Linux Kernel release v4.0 and evaluated \rev{on all components of} v4.1). 
The components of the following release are converted into sequences that are input to the trained model to get the output sequences.
Then, \our compares the output sequences generated by the trained model with the input sequences.
A component is considered vulnerable if any of the output sequences differ from the input sequences, otherwise considered as non-vulnerable. 
This training-testing process is repeated for all available releases.

\rev{For our release-based experiments where we train the models of different approaches on one release and test the trained models on the next release, in total we have 36 releases of Linux Kernel, 10 releases of Wireshark, and 10 releases of OpenSSL, as mentioned in Table~\ref{tab:datasetstatistics}. In case of \textit{(n)} releases available to us for a system, we can only perform \textit{(n--1)} experiments because in chronological order, the last experiment would be to train a model on \textit{(n--1)$^{th}$} release and test the trained model on \textit{(n)$^{th}$} release. The reason for such is that we do not have a release to test a model trained on the \textit{nth} release. Hence, for 1 approach, we performed 35 experiments for Linux Kernel, 9 experiments for Wireshark, and 9 experiments for Wireshark. That results to 53 experiments in total (35 + 9 + 9 = 53), for 1 approach.}

\textit{\rev{\realset} - Used in RQ4: }
In contrast to the \emph{clean} training \rev{data} settings, in \rev{\realset} we consider the date when the vulnerability was fixed.
Vulnerability fixing date determines whether a vulnerability is included in the training dataset or not.
\rev{In these settings,} a prediction model (for \rev{one} release \rev{of the system}) is trained only \rev{on} those vulnerabilities that were fixed before the next release date.
Then, the trained model is evaluated \rev{on all the components of the} following release of the same system.

\subsection{Benchmarks for Vulnerability Prediction}
\label{subsec:benchmarks}
To assess effectiveness, we compare \our with 
existing vulnerability prediction techniques\rev{. To perform the comparison we use \emph{FrameVPM}, a framework enabling the replication and comparison of vulnerability prediction approaches, introduced by Jimenez~\etal\cite{Jimenez+2019}. Overall, we compare \our with}:

\sm: Complexity metrics have been extensively used for defect prediction (e.g. \cite{Hall+2012}) and vulnerability prediction (e.g. \cite{10.1145/1414004.1414065,ShinMWO2011,10.1016/j.sysarc.2010.06.003,TheisenW20}).
It is based on the idea that complex code is difficult to maintain and test, and thus has a higher chance of having vulnerabilities than simple code.
\rev{Using FrameVPM, we replicate and compare with the original study from Shin \etal\cite{ShinMWO2011} that rely on features related to following metrics: 
\begin{enumerate}
\item Complexity and Coupling
\begin{enumerate}[label=(\alph*)]
\item \emph{LinesOfCode}: lines of code; 
\item \emph{PreprocessorLines}: preprocessing lines of code; 
\item \emph{CommentDensity ratio}: lines of comments to lines of code; 
\item \emph{CountDeclFunction}: number of functions defined; 
\item \emph{CountDeclVariable}: number of variables defined; 
\item \emph{CC}(sum, avg, max): sum, average and max cyclomatic complexity; 
\item \emph{SCC}(sum, avg, max): strict cyclomatic complexity\cite{ShinMWO2011}; 
\item \emph{CCE}(sum, avg, max): essential cyclomatic complexity\cite{ShinMWO2011}; 
\item \emph{MaxNesting}(sum, avg, max): maximum nesting level of control constructs; 
\item \emph{fanIn}(sum, avg, max): number of inputs, i.e., input parameters and global variables to functions;
\item \emph{fanOut}(sum, avg, max): number of outputs, i.e., assignments to global variables and parameters of function calls.
\end{enumerate}
\item Code Churn: \emph{added lines, modified lines and deleted lines in the history of a component.}
\item Developer Activity Metrics: 
\begin{enumerate}[label=(\alph*)]
\item \emph{number of commits impacting a component;}
\item \emph{number of developers modified a component;}
\item \emph{current number of developers working on a component.}
\end{enumerate}
\end{enumerate}
}

\norev{
\tm: It considers a source code component as a collection of terms associated with frequencies, also known as \emph{Bag of Words} (BoW), used for vulnerability prediction \cite{6860243}.
The source code is broken into a vector of code tokens, and the frequency of each token is then used as the features to build the vulnerability prediction model.
Further improvements have been performed to significantly improve its performance, e.g., by pooling frequency values in different bins according to particular criteria to discretize BoW's features~\cite{6860243,10.5555/1643031.1643034,TheisenW20}.
}

\im and \fc: The work of \emph{Neuhaus \etal}~\cite{Neuhaus+2007} is based on the observation that the vulnerable components tend to import and call a particular small set of functions.
Thus, the features of this simple prediction model are the components’ imports and function calls.
Following the suggestions of \emph{FrameVPM}, we use imports and function calls as separate sets of features.
We train one model based on \rev{\im} and another based on \rev{\fc}, thus implementing one model per set of features.

\rev{

\dev: The work of \emph{Zhou \etal}~\cite{zhou2019devign} emphasizes the use of graph neural network for vulnerability detection. With Abstract Syntax Tree (AST) as the backbone, \emph{Zhou \etal} proposed to convert components (vulnerable/non-vulnerable) as code property graphs which helps to solve the problem of information loss during learning. To perform component classification, (\ie graph-level classification), graph neural network models are trained which are composed of gated graph recurrent layer and convolutional layer, that enables to learn the vulnerable programming pattern. Since the authors' implementation of the approach is not available, we implemented \dev based on our understanding of~\cite{zhou2019devign} and made it publicly available\footnote{\rev{https://github.com/garghub/TROVON/tree/main/devign}}.

\lstm and \lstmrf: The work of \emph{Dam \etal}~\cite{Dam+2018} focuses to capture \emph{semantic} features of code components (vulnerable/non-vulnerable) and using these features to perform vulnerability prediction. \emph{Dam \etal} asserted that \emph{Long Short Term Memory}~(LSTM)~\cite{HochreiterHochreiter1997} is highly effective in learning long-term dependencies in sequential data such as text and speech, and can be used to learn features that represent both the semantics of code tokens (semantic features) and the sequential structure of source code (syntactic features). In this approach, components are encoded using the embedding layer, and along with labels (vulnerable/non-vulnerable), are used to train LSTM models. Although these trained LSTM models are capable of prediction, \ie to provide a probability of a component being vulnerable, the approach extends a step further. The embeddings for the components are extracted using the trained LSTM models, and are used to train binary classifier. Finally, the trained binary classifier provides the probability/likelihood of a component being vulnerable. 
For \lstm approach, we used the trained LSTM models for predictions, and for \lstmrf approach, we used trained binary classifiers for predictions. Here as well, due to unavailable authors' implementation, we implemented the approach based on our understanding of~\cite{HochreiterHochreiter1997} and made it publicly available\footnote{\rev{https://github.com/garghub/TROVON/tree/main/lstm-rf}}.
}

\subsection{Performance measurement}
\label{subsec:performancemeasurement}
Vulnerability prediction modeling is a binary classification problem, thus it can result in four types of outputs:
Given a vulnerable component, if it is predicted as vulnerable, then it is a true positive (TP); otherwise, it is a false negative (FN).
Given a non-vulnerable component, if it is predicted as non-vulnerable, then it is a true negative (TN); otherwise, it is a false positive (FP).
From these, we can compute the traditional evaluation metrics such as \emph{Precision}, \emph{Recall}, and \emph{F-measure} scores, which quantitatively evaluate the prediction accuracy of vulnerability prediction models.
\begin{align*}
   & \emph{Precision} = \frac{TP}{TP + FP}  \hspace{1em} \emph{Recall} = \frac{TP}{TP + FN} 
   \hspace{1em}\emph{F-measure} = \frac{2 \times \emph{Precision} \times \emph{Recall}}{\emph{Precision} + \emph{Recall}}
\end{align*}

Intuitively, \emph{Precision} indicates the ratio of correctly predicted positives over all the considered positives.
\emph{Recall} indicates the ratio of correctly predicted positives over all the actual positives.
\emph{F-measure} indicates the weighted harmonic mean of Precision and Recall.

Yet, these metrics do not take into account the true negatives and can be misleading, especially in the case of imbalanced data.
Hence, we complement these with the \emph{Matthews Correlation Coefficient (MCC)} \cite{MATTHEWS1975442}, a reliable metric of the quality of prediction models \cite{6824804}.
It is generally regarded as a balanced measure that can be used even when the classes are of very different sizes, e.g. in the case of Linux Kernel, 3\% vulnerable components (positives) over 97\% non-vulnerable components (negatives).
\emph{MCC} is calculated as:
\[\emph{MCC} = \frac{TP \times TN - FP \times FN}{\sqrt{(TP + FP)(TP + FN)(TN + FP)(TN + FN)}}\]
\emph{MCC} returns a coefficient between 1 and -1.
An MCC value of 1 indicates a perfect prediction, while a value of -1 indicates a perfect inverse prediction \ie a total disagreement between prediction and reality.
MCC value of 0 indicates that the prediction performance is equivalent to random guessing.

\section{Experimental Results}
\label{subsec:results}

\begin{table}[]
  \begin{center}
    \caption{Prediction with clean training data, \rev{aka \cleanset} (RQ1)}
    \label{tab:rq1results}
    \resizebox{0.68\textwidth}{!}{
    \begin{tabular}{|l|r|r|r|r|r|}
    \hline
      \rule{0pt}{3ex}\makecell{\textbf{Release}}            & \textbf{MCC} & \textbf{F-measure} & \textbf{Precision} & \textbf{Recall} & \textbf{\rev{Total Vuln. Comp.}} \\
      \hline
      \multicolumn{6}{|c|}{\rule{0pt}{3ex}Linux Kernel}                                                                       \\
      \hline
         \rule{0pt}{3ex}v3.0 & \rev{0.70} & \rev{0.86} & \rev{0.84} & \rev{0.89} & \rev{598} \\
            v3.1 & \rev{0.72} & \rev{0.87} & \rev{0.82} & \rev{0.92} & \rev{612} \\
            v3.2 & \rev{0.75} & \rev{0.88} & \rev{0.86} & \rev{0.91} & \rev{612} \\
            v3.3 & \rev{0.70} & \rev{0.86} & \rev{0.82} & \rev{0.91} & \rev{609} \\
            v3.4 & \rev{0.73} & \rev{0.88} & \rev{0.84} & \rev{0.91} & \rev{607} \\
            v3.5 & \rev{0.72} & \rev{0.86} & \rev{0.94} & \rev{0.79} & \rev{609} \\
            v3.6 & \rev{0.74} & \rev{0.88} & \rev{0.86} & \rev{0.90} & \rev{640} \\
            v3.7 & \rev{0.67} & \rev{0.85} & \rev{0.82} & \rev{0.89} & \rev{640} \\
            v3.8 & \rev{0.78} & \rev{0.89} & \rev{0.92} & \rev{0.87} & \rev{632} \\
            v3.9 & \rev{0.69} & \rev{0.86} & \rev{0.83} & \rev{0.90} & \rev{633} \\
            v3.10 & \rev{0.77} & \rev{0.89} & \rev{0.88} & \rev{0.90} & \rev{637} \\
            v3.11 & \rev{0.85} & \rev{0.93} & \rev{0.93} & \rev{0.92} & \rev{613} \\
            v3.12 & \rev{0.76} & \rev{0.89} & \rev{0.88} & \rev{0.90} & \rev{584} \\
            v3.13 & \rev{0.72} & \rev{0.87} & \rev{0.82} & \rev{0.92} & \rev{578} \\
            v3.14 & \rev{0.85} & \rev{0.93} & \rev{0.93} & \rev{0.93} & \rev{573} \\
            v3.15 & \rev{0.78} & \rev{0.89} & \rev{0.89} & \rev{0.90} & \rev{554} \\
            v3.16 & \rev{0.80} & \rev{0.91} & \rev{0.92} & \rev{0.89} & \rev{553} \\
            v3.17 & \rev{0.81} & \rev{0.91} & \rev{0.91} & \rev{0.91} & \rev{443} \\
            v3.18 & \rev{0.81} & \rev{0.91} & \rev{0.93} & \rev{0.89} & \rev{428} \\
            v3.19 & \rev{0.72} & \rev{0.87} & \rev{0.84} & \rev{0.91} & \rev{420} \\
            v4.0 & \rev{0.88} & \rev{0.94} & \rev{0.96} & \rev{0.92} & \rev{417} \\
            v4.1 & \rev{0.86} & \rev{0.93} & \rev{0.94} & \rev{0.93} & \rev{417} \\
            v4.2 & \rev{0.77} & \rev{0.88} & \rev{0.96} & \rev{0.82} & \rev{410} \\
            v4.3 & \rev{0.84} & \rev{0.92} & \rev{0.94} & \rev{0.90} & \rev{391} \\
            v4.4 & \rev{0.82} & \rev{0.92} & \rev{0.91} & \rev{0.93} & \rev{371} \\
            v4.5 & \rev{0.79} & \rev{0.90} & \rev{0.92} & \rev{0.88} & \rev{347} \\
            v4.6 & \rev{0.79} & \rev{0.90} & \rev{0.88} & \rev{0.93} & \rev{330} \\
            v4.7 & \rev{0.79} & \rev{0.90} & \rev{0.91} & \rev{0.90} & \rev{310} \\
            v4.8 & \rev{0.83} & \rev{0.92} & \rev{0.91} & \rev{0.92} & \rev{284} \\
            v4.9 & \rev{0.80} & \rev{0.90} & \rev{0.90} & \rev{0.90} & \rev{259} \\
            v4.10 & \rev{0.79} & \rev{0.90} & \rev{0.92} & \rev{0.88} & \rev{233} \\
            v4.11 & \rev{0.75} & \rev{0.88} & \rev{0.87} & \rev{0.90} & \rev{194} \\
            v4.12 & \rev{0.78} & \rev{0.89} & \rev{0.93} & \rev{0.86} & \rev{176} \\
            v4.13 & \rev{0.79} & \rev{0.90} & \rev{0.94} & \rev{0.86} & \rev{133} \\
            v4.14 & \rev{0.80} & \rev{0.91} & \rev{0.91} & \rev{0.90} & \rev{113} \\
     \hline
     \multicolumn{6}{|c|}{\rule{0pt}{3ex}Wireshark} \\
     \hline
         \rule{0pt}{3ex}v1.8.0 & \rev{0.50} & \rev{0.69} & \rev{0.97} & \rev{0.53} & \rev{138} \\
            v1.10.0 & \rev{0.58} & \rev{0.77} & \rev{0.92} & \rev{0.67} & \rev{168} \\
            v1.11.0 & \rev{0.78} & \rev{0.88} & \rev{0.97} & \rev{0.81} & \rev{168} \\
            v1.12.0 & \rev{0.58} & \rev{0.76} & \rev{0.95} & \rev{0.63} & \rev{165} \\
            v1.99.0 & \rev{0.71} & \rev{0.85} & \rev{0.95} & \rev{0.77} & \rev{156} \\
            v2.0.0 & \rev{0.59} & \rev{0.78} & \rev{0.93} & \rev{0.67} & \rev{123} \\
            v2.1.0 & \rev{0.74} & \rev{0.86} & \rev{0.98} & \rev{0.76} & \rev{116} \\
            v2.2.0 & \rev{0.67} & \rev{0.83} & \rev{0.93} & \rev{0.75} & \rev{93} \\
            v2.4.0 & \rev{0.17} & \rev{0.65} & \rev{0.69} & \rev{0.61} & \rev{79} \\
     \hline
     \multicolumn{6}{|c|}{\rule{0pt}{3ex}OpenSSL} \\
     \hline
         \rule{0pt}{3ex}v0.9.3 & \rev{0.83} & \rev{0.91} & \rev{1.00} & \rev{0.83} & \rev{53} \\
            v0.9.4 & \rev{0.83} & \rev{0.91} & \rev{1.00} & \rev{0.83} & \rev{56} \\
            v0.9.5 & \rev{0.83} & \rev{0.91} & \rev{1.00} & \rev{0.83} & \rev{56} \\
            v0.9.6 & \rev{0.67} & \rev{0.80} & \rev{1.00} & \rev{0.67} & \rev{65} \\
            v0.9.7 & \rev{0.71} & \rev{0.83} & \rev{1.00} & \rev{0.71} & \rev{78} \\
            v0.9.8 & \rev{0.71} & \rev{0.83} & \rev{1.00} & \rev{0.71} & \rev{75} \\
            v1.0.0 & \rev{0.71} & \rev{0.84} & \rev{0.96} & \rev{0.75} & \rev{71} \\
            v1.0.1 & \rev{0.73} & \rev{0.87} & \rev{0.91} & \rev{0.82} & \rev{48} \\
            v1.0.2 & \rev{0.67} & \rev{0.80} & \rev{1.00} & \rev{0.67} & \rev{26} \\
     \hline
     \multicolumn{6}{|c|}{\rule{0pt}{3ex}Overall} \\
     \hline
     \rule{0pt}{3ex}Average & \rev{0.74} & \rev{0.87} & \rev{0.91} & \rev{0.84} & \rev{334}\\
     \hline
     \rule{0pt}{3ex}Median & \rev{0.76} & \rev{0.88} & \rev{0.92} & \rev{0.89} & \rev{330}\\
     \hline
     \end{tabular}
     }
 \end{center}
\end{table}
\vspace{-0.5em}

\subsection{Prediction with clean training data, \rev{aka \cleanset} (RQ1)}

\label{subsubsec:rq1}
Table~\ref{tab:rq1results} records the prediction performance results for the experiments conducted on the 56 releases we study, \ie 36 releases of Linux Kernel, 10 of Wireshark, and 10 of OpenSSL\rev{, and the total number of vulnerable components present in every release}.
As mentioned earlier, here the model is trained on a release and evaluated against the following (next) release of the same system.
\our obtained an overall average (and median) of \emph{MCC=~\rev{0.74}} (\rev{0.76}), \emph{F-measure=~\rev{0.87}} (\rev{0.88}), \emph{Precision=~\rev{0.91}} (\rev{0.92}), and \emph{Recall=~\rev{0.84}} (\rev{0.89}) in prediction of vulnerable components in the next release of a project.
\rev{For} almost all \rev{releases}, \our's prediction models trained with the clean data achieved above 0.65 MCC (49 out of 53 \rev{releases}), above 0.75 F-measure (51 out of 53 \rev{releases}), above 0.80 Precision (52 out of 53 \rev{releases}), and above 0.70 Recall (49 out of 53 \rev{releases}).
The results achieved by \our indicate that the suggested predictions can be considered actionable for security engineers looking to prioritize security inspection and testing efforts~\cite{shin_can_2013}.\\

\vspace{-0.5em}
\begin{tcolorbox}
  Answer to RQ1: The vulnerability prediction models built on \our successfully predict the vulnerable components with an average MCC score of \rev{0.74}, which can be considered actionable for security engineers to prioritize components for security inspection.
\end{tcolorbox}
\vspace{-0.5em}

\begin{figure*}[t]
  \begin{center}
    \makebox[\textwidth]{\includegraphics[width=\textwidth]{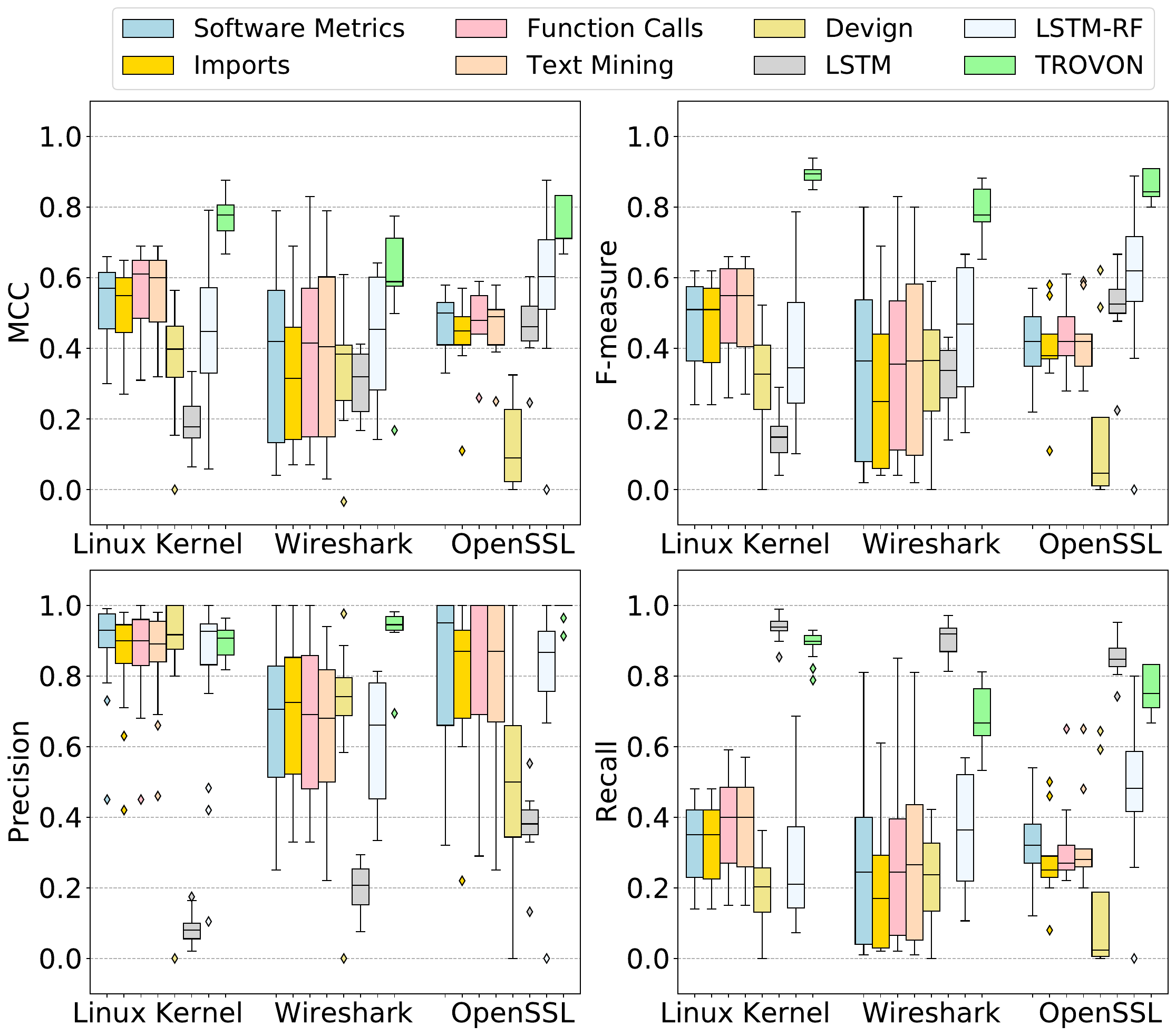}}
\caption{Comparison with existing approaches (RQ2) \rev{in \cleanset}: When trained with clean data, \our outperforms existing approaches with an average improvement in MCC, F-measure, Precision, and Recall of \rev{80.39\%}, \rev{132.95\%}, \rev{83.96\%}, and \rev{155.33\%}, respectively.}
    \label{fig:filepredictionidealbarplot}
  \end{center}
\end{figure*}

\begin{table}[t]
  \begin{center}
    \caption{(RQ2) Comparison between existing techniques and \our~\rev{under \cleanset} - average (\rev{and} median)}
    \label{tab:rq2results}
    \begin{tabular}{|l|r|r|r|r|}
\hline
\rule{0pt}{3ex}\textbf{Approach} & \textbf{MCC}  & \textbf{F-measure}  & \textbf{Precision}  & \textbf{Recall} \\ \hline
\rule{0pt}{3ex}\sm & \rev{0.49} \rev{(0.53)} & \rev{0.44} \rev{(0.48)} & \rev{0.85} \rev{(0.92)} & \rev{0.32} \rev{(0.34)} \\ \hline
\rule{0pt}{3ex}\im & \rev{0.46} \rev{(0.49)} & \rev{0.43} \rev{(0.44)} & \rev{0.83} \rev{(0.88)} & \rev{0.30} \rev{(0.29)} \\ \hline
\rule{0pt}{3ex}\fc & \rev{0.52} \rev{(0.56)} & \rev{0.48} \rev{(0.50)} & \rev{0.84} \rev{(0.89)} & \rev{0.36} \rev{(0.35)} \\ \hline
\rule{0pt}{3ex}\tm & \rev{0.52} \rev{(0.55)} & \rev{0.48} \rev{(0.51)} & \rev{0.83} \rev{(0.88)} & \rev{0.36} \rev{(0.38)} \\ \hline
\rule{0pt}{3ex}\rev{\dev} & \rev{0.33 (0.36)} & \rev{0.29 (0.32)} & \rev{0.79 (0.89)} & \rev{0.19 (0.19)} \\ \hline
\rule{0pt}{3ex}\rev{\lstm} & \rev{0.25 (0.22)} & \rev{0.23 (0.18)} & \rev{0.15 (0.09)} & \bf{\rev{0.92 (0.93)}} \\ \hline
\rule{0pt}{3ex}\rev{\lstmrf} & \rev{0.47 (0.49)} & \rev{0.43 (0.42)} & \rev{0.80 (0.89)} & \rev{0.32 (0.29)} \\ \hline
\rule{0pt}{3ex}\our & \bf{\rev{0.74} \rev{(0.76)}} & \bf{\rev{0.87} \rev{(0.88)}} & \bf{\rev{0.91} \rev{(0.92)}} & \rev{0.84} \rev{(0.89)} \\ \hline
\end{tabular}
  \end{center}
\end{table}

\subsection{Comparison with existing techniques (RQ2)}
\label{subsubsec:rq2}
Figure \ref{fig:filepredictionidealbarplot} shows the performance comparison of \our with existing approaches in a box plot format.
Box plots show the distribution of performance indicators (MCC, F-measure, Precision, Recall) for the techniques per project.

We can observe that \our outperforms the others by achieving higher MCC scores.
Table \ref{tab:rq2results} summarizes the overall performance of the techniques.
Interestingly, \our achieved higher prediction performance in comparison to existing techniques, with a statistically significant\footnote{We compared the MCC values by using Wilcoxon sign-rank-test~\cite{Wilcoxon1945}, and obtained a $\mathit{p-value} < \rev{6.2}\mathrm{\rev{e}}{\rev{-9}}$ \rev{with existing approaches.}
We also compared the effect size of MCC values, by using the Vargha-Delaney A measure~\cite{VarghaDelaney2000}, and obtained a value of \rev{lower than $0.07$ in every case}, clearly indicating that \our significantly outperforms existing techniques.} difference.
We can also observe that the technique \fc outperforms the others ( \sm, \im, \tm, \rev{\dev, \lstm, and \lstmrf} ) with its average MCC of \rev{0.52}. 
\our even outperforms \fc with its \rev{40.84\%} higher MCC and \rev{80.67\%} higher F-measure.
It is worth mentioning that the average improvement offered by \our is \rev{8.68\%} in Precision and \rev{134.73\%} in Recall, in comparison to \fc.\\
The results show that \our can provide comparatively better guidance to security engineers than existing techniques, to prioritize components for security inspection~\cite{shin_can_2013}.\\

\vspace{-0.5em}
\begin{tcolorbox}
  Answer to RQ2: When trained with clean data, \our has significantly higher prediction ability, \ie~\rev{on average, 80.39\%} improvement in MCC score than existing approaches, which shows that \our can guide security engineers comparatively better than existing techniques to prioritize security inspection and testing efforts.
\end{tcolorbox}
\vspace{-0.5em}

\subsection{Predictions on Seen vs Unseen Vulnerable Components (RQ3)}
\label{subsubsec:rq3}

Table~\ref{tab:rq3results01} shows the average percentages of the seen vulnerable \rev{components} correctly predicted by \our and existing techniques across 56 releases of the systems.
On average, the models that are based on \our predict \rev{92.79\%}, \rev{69.48\%} and \rev{87.19\%} of the seen vulnerable components in Linux Kernel, Wireshark, and OpenSSL project releases, respectively.
\rev{The models based on \lstm performs the best in identifying already seen vulnerable components, \ie 96.69\%, 76.43\%, and 95.77\% of the vulnerable components identified correctly in Linux Kernel, Wireshark, and OpenSSL project releases, respectively.}
The percentages gained by \our are higher than existing techniques, \rev{except \lstm,} by \rev{44.12\%} for Linux Kernel releases, \rev{17.19\%} for Wireshark releases, and \rev{33.81\%} for OpenSSL releases, indicating a \rev{high} learning potential.

\begin{table}[t]
  \begin{center}
    \caption{(RQ3) Comparison between existing techniques and \our~\rev{wrt to their ability to predict correctly already} seen vulnerable components\rev{, i.e., (classify then  as vulnerable)}
    }
    \label{tab:rq3results01}
    \begin{tabular}{|l|r|r|r|}
    \hline
 \rule{0pt}{4ex}{\textbf{Approach}} & \makecell{\textbf{Linux Kernel}\\36 releases}  & \makecell{\textbf{Wireshark}\\10 releases}     & \makecell{\textbf{OpenSSL}\\10 releases}       \\
      \hline
      \rule{0pt}{3ex}\sm & 48.12\% & 54.84\% & 54.17\% \\
      \hline
      \rule{0pt}{3ex}\im & 48.12\% & 60.76\% & 50.00\% \\
      \hline
      \rule{0pt}{3ex}\fc & 58.65\% & 52.69\% & 64.58\% \\
      \hline
      \rule{0pt}{3ex}\tm & 57.14\% & 56.99\% & 64.58\% \\
      \hline
      \rule{0pt}{3ex}\rev{\dev} & \rev{32.34\%} & \rev{39.64\%} & \rev{35.69\%} \\
      \hline
      \rule{0pt}{3ex}\rev{\lstm} & \bf{\rev{96.69\%}} & \rev{\bf{76.43\%}} & \rev{\bf{95.77\%}} \\
      \hline
      \rule{0pt}{3ex}\rev{\lstmrf} & \rev{47.66\%} & \rev{48.81\%} & \rev{51.25\%} \\
      \hline
      \rule{0pt}{3ex}\our & 92.79\% & 69.48\% & 87.19\% \\      \hline
    \end{tabular}
  \end{center}
\end{table}

\begin{table}[t]
  \begin{center}
    \caption{(RQ3) Comparison between existing techniques and \our~\rev{wrt to their ability to predict correctly already} unseen vulnerable components\rev{, i.e., (classify then  as vulnerable)}
    }
    \label{tab:rq3results02}
    \begin{tabular}{|l|r|r|r|}
          \hline
      \rule{0pt}{4ex}\textbf{Approach} & \makecell{\textbf{Linux Kernel}\\36 releases}  & \makecell{\textbf{Wireshark}\\10 releases}     & \makecell{\textbf{OpenSSL}\\10 releases} \\
      \hline
\rule{0pt}{3ex}\sm  & 09.09\% & 15.48\% & 18.18\% \\
      \hline
\rule{0pt}{3ex}\im & 50.00\% & 08.93\% & 23.08\%\\
      \hline
\rule{0pt}{3ex}\fc & 56.10\% & 60.00\% & 09.09\% \\
      \hline
\rule{0pt}{3ex}\tm & 45.45\% & 16.07\% & 18.18\% \\
      \hline
\rule{0pt}{3ex}\rev{\dev} & \rev{32.54\%} & \rev{33.13\%} & \rev{14.99\%} \\
      \hline
\rule{0pt}{3ex}\rev{\lstm} & \rev{25.79\%} & \rev{27.63\%} & \rev{23.02\%} \\
      \hline
\rule{0pt}{3ex}\rev{\lstmrf} & \rev{36.39\%} & \rev{25.62\%} & \rev{18.01\%} \\
      \hline
\rule{0pt}{3ex}\our & \bf{76.53\%} & \bf{91.03\%} & \bf{60.07\%} \\
 \hline
    \end{tabular}
  \end{center}
\end{table}

Table~\ref{tab:rq3results02} shows the average percentages of the unseen vulnerable component prediction. 
On average, \our based trained models predict \rev{76.53\%}, \rev{91.03\%} and \rev{60.07\%} of the unseen vulnerable components in Linux Kernel, Wireshark, and OpenSSL project releases, respectively. 
The percentages gained by \our are higher than existing techniques by \rev{40.05\%} for Linux Kernel releases, \rev{64.34\%} for Wireshark releases, and \rev{42.28\%} for OpenSSL releases, reflecting higher generalization capability in comparison to existing techniques.
It is worth noting that \our obtains all the above mentioned percentages with an MCC of \rev{0.74}, \rev{on average}, which is \rev{80.39\%} higher than existing techniques.

\begin{tcolorbox}
  Answer to RQ3: \rev{The models trained on \our have higher learning potential and generalization capability in comparison to existing approaches in almost all cases.}
\end{tcolorbox}
\vspace{-1em}

\subsection{Comparison with existing techniques under \rev{\realset} (RQ4)}
\label{subsubsec:rq4} 
\begin{figure*}[t]
  \begin{center}
    \makebox[\textwidth]{\includegraphics[width=\textwidth]{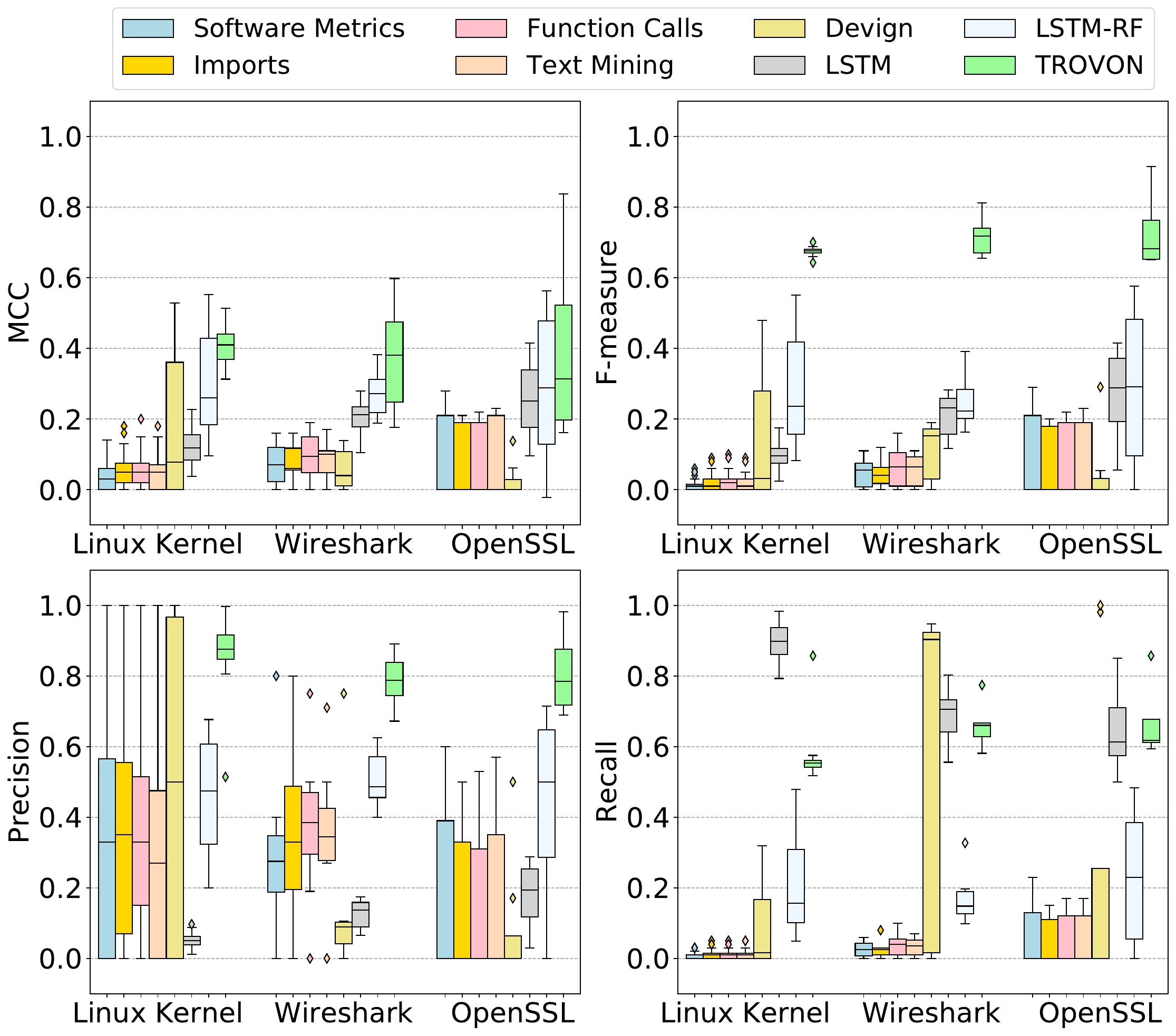}}
    \caption{Comparison with existing techniques in \rev{\realset} (RQ4): Despite a reduced performance \rev{of models when trained with realistic training data}, \our significantly outperforms existing techniques with \rev{3.63} times higher MCC, \rev{11.82} times higher F-measure, \rev{2.66} times higher Precision, and \rev{15.25} times higher Recall, respectively.
    }
    \label{fig:filepredictionrealisticbarplot}
  \end{center}
\vspace{-1em}
\end{figure*}

As mentioned before, in \rev{\realset}, a model is trained only on the vulnerabilities of a release that were detected / made public before the next release date of the system.
This unavoidably introduces mislabeling noise because every component that has no vulnerabilities uncovered before the next release date, is considered non-vulnerable during training.
Figure \ref{fig:filepredictionrealisticbarplot} shows that the performance of all the techniques is considerably reduced in the \rev{\realset}, in comparison to the \cleanset. 
The results are in accordance with Jimenez \etal~\cite{Jimenez+2019}. 
Despite this drop in performance, \our outperforms existing techniques with a \rev{statistically significant}\footnote{We compared the MCC values using Wilcoxon sign-rank-test and obtained a $\mathit{p-value} < \rev{7.7}\mathrm{\rev{e}}{\rev{-9}}$ \rev{with existing approaches.} We also compared the MCC values with the Vargha-Delaney A measure and obtained a value lower than \rev{$0.03$ in every case}, indicating that \our significantly outperforms existing techniques.} sizeable difference.

\begin{table}[t]
  \begin{center}
    \caption{(RQ4) Comparison between existing techniques and \our~\rev{under \realset} - average (median)}
    \label{tab:rq4results}
        \begin{tabular}{|l|r|r|r|r|}
\hline
 \rule{0pt}{3ex}\textbf{Approach} & \textbf{MCC} & \textbf{F-measure} & \textbf{Precision} & \textbf{Recall} \\ \hline
\rule{0pt}{3ex}\sm & \rev{0.06} \rev{(0.03)} & \rev{0.03} \rev{(0.01)} & \rev{0.31} \rev{(0.30)} & \rev{0.02} \rev{(0.01)} \\ \hline
\rule{0pt}{3ex}\im  & \rev{0.06} \rev{(0.06)} & \rev{0.04} \rev{(0.02)} & \rev{0.34} \rev{(0.33)} & \rev{0.02} \rev{(0.01)} \\ \hline
\rule{0pt}{3ex}\fc & \rev{0.07} \rev{(0.05)} & \rev{0.04} \rev{(0.02)} & \rev{0.34} \rev{(0.33)} & \rev{0.03} \rev{(0.01)} \\ \hline
\rule{0pt}{3ex}\tm & \rev{0.06} \rev{(0.05)} & \rev{0.04} \rev{(0.01)} & \rev{0.29} \rev{(0.28)} & \rev{0.02} \rev{(0.01)} \\ \hline
\rule{0pt}{3ex}\rev{\dev} & \rev{0.13 (0.02)} & \rev{0.12 (0.03)} & \rev{0.34 (0.06)} & \rev{0.18 (0.02)} \\ \hline
\rule{0pt}{3ex}\rev{\lstm} & \rev{0.16 (0.14)} & \rev{0.14 (0.11)} & \rev{0.08 (0.06)} & \bf{\rev{0.83 (0.86)}} \\ \hline
\rule{0pt}{3ex}\rev{\lstmrf} & \rev{0.29 (0.27)} & \rev{0.28 (0.23)} & \rev{0.47 (0.49)} & \rev{0.21 (0.15)} \\ \hline
\rule{0pt}{3ex}\our & \bf{\rev{0.39} \rev{(0.41)}} & \bf{\rev{0.69} \rev{(0.68)}} & \bf{\rev{0.86} \rev{(0.87)}} & \rev{0.58} \rev{(0.56)} \\ \hline
    \end{tabular}
  \end{center}
\end{table}

Table \ref{tab:rq4results} shows the overall average and median performance statistics for each technique. 
We can observe that the technique \rev{\lstmrf} outperforms the other existing techniques (\sm, \im, \rev{\fc, \tm, \dev, and \lstm}) with its average MCC of \rev{0.29}.
\our even outperforms \lstmrf in all the performance measures, \rev{\ie 35.52\% higher MCC, 148.91\% higher F-measure, 81.61\% higher Precision, and 183.90\% higher Recall, in comparison to \lstmrf}.
This indicates that \our has much higher accuracy in vulnerability prediction than existing techniques in the \rev{\realset} as well.\\

\begin{tcolorbox}
  Answer to RQ4: Under the \rev{\realset}, \our based models obtain significantly higher accuracy in vulnerability prediction, \ie~\rev{3.63} times higher MCC scores than existing techniques.
\end{tcolorbox}

\minrev{
\section{TROVON with Bi-LSTM}
\label{sec:discussion}
Although training a machine translator (viz. an encoder-decoder sequence to sequence model) to identify vulnerable components, is an integral part of \our's architecture, we also replicated our experiments with \bilstm models. We kept the entire experimental setting the same, (i.e., both \cleanset and \realset with the corresponding training and test sets) and trained \bilstm models instead of training sequence to sequence models. For this experiment, we adhere to the key idea of \our and train \bilstm models on the validated data, (i.e., only on components known to be vulnerable and leave aside the non-vulnerable ones). We name this approach \ourbilstm.

\begin{table}[t]
  \begin{center}
    \caption{\minrev{Comparison between \ourbilstm and \our under \cleanset~- average (median)}}
    \label{tab:discussion_clean_results}
        \begin{tabular}{|l|r|r|r|r|}
        \hline
        \rule{0pt}{3ex}\textbf{\minrev{Approach}} & \textbf{\minrev{MCC}}  & \textbf{\minrev{F-measure}}  & \textbf{\minrev{Precision}}  & \textbf{\minrev{Recall}} \\ \hline
        \multicolumn{5}{|c|}{\rule{0pt}{3ex}\minrev{Linux Kernel}} \\ \hline
        \rule{0pt}{3ex}\minrev{\ourbilstm} & \minrev{0.73 (0.70)} & \minrev{0.84 (0.83)} & \minrev{0.84 (0.83)} & \minrev{0.84 (0.84)} \\ \hline
        \rule{0pt}{3ex}\minrev{\our} & \bf{\minrev{0.78 (0.78)}} & \bf{\minrev{0.89 (0.89)}} & \bf{\minrev{0.89 (0.91)}} & \bf{\minrev{0.90 (0.90)}} \\ \hline
        \multicolumn{5}{|c|}{\rule{0pt}{3ex}\minrev{Wireshark}} \\ \hline
        \rule{0pt}{3ex}\minrev{\ourbilstm} & \minrev{0.54 (0.54)} & \minrev{0.72 (0.72)} & \minrev{0.85 (0.85)} & \minrev{0.63 (0.61)} \\ \hline
        \rule{0pt}{3ex}\minrev{\our} & \bf{\minrev{0.59 (0.59)}} & \bf{\minrev{0.79 (0.78)}} & \bf{\minrev{0.92 (0.95)}} & \bf{\minrev{0.69 (0.67)}} \\ \hline
        \multicolumn{5}{|c|}{\rule{0pt}{3ex}\minrev{OpenSSL}} \\ \hline
        \rule{0pt}{3ex}\minrev{\ourbilstm} & \minrev{0.71 (0.68)} & \minrev{0.82 (0.79)} & \minrev{0.93 (0.98)} & \minrev{0.73 (0.68)} \\ \hline
        \rule{0pt}{3ex}\minrev{\our} & \bf{\minrev{0.74 (0.71)}} & \bf{\minrev{0.86 (0.84)}} & \bf{\minrev{0.99 (0.99)}} & \bf{\minrev{0.76 (0.75)}} \\ \hline
        \end{tabular}
  \end{center}
\end{table}

\begin{table}[t]
  \begin{center}
    \caption{\minrev{Comparison between \ourbilstm and \our under \realset~- average (median)}}
    \label{tab:discussion_realistic_results}
        \begin{tabular}{|l|r|r|r|r|}
        \hline
        \rule{0pt}{3ex}\textbf{\minrev{Approach}} & \textbf{\minrev{MCC}}  & \textbf{\minrev{F-measure}}  & \textbf{\minrev{Precision}}  & \textbf{\minrev{Recall}} \\ \hline
        \multicolumn{5}{|c|}{\rule{0pt}{3ex}\minrev{Linux Kernel}} \\ \hline
        \rule{0pt}{3ex}\minrev{\ourbilstm} & \minrev{0.38 (0.39)} & \minrev{0.65 (0.67)} & \minrev{0.84 (0.87)} & \minrev{0.53 (0.54)} \\ \hline
        \rule{0pt}{3ex}\minrev{\our} & \bf{\minrev{0.40 (0.41)}} & \bf{\minrev{0.68 (0.68)}} & \bf{\minrev{0.88 (0.88)}} & \bf{\minrev{0.56 (0.55)}} \\ \hline
        \multicolumn{5}{|c|}{\rule{0pt}{3ex}\minrev{Wireshark}} \\ \hline
        \rule{0pt}{3ex}\minrev{\ourbilstm} & \minrev{0.34 (0.36)} & \minrev{0.66 (0.65)} & \minrev{0.73 (0.70)} & \minrev{0.61 (0.62)} \\ \hline
        \rule{0pt}{3ex}\minrev{\our} & \bf{\minrev{0.37 (0.38)}} & \bf{\minrev{0.72 (0.72)}} & \bf{\minrev{0.79 (0.79)}} & \bf{\minrev{0.66 (0.66)}} \\ \hline
        \multicolumn{5}{|c|}{\rule{0pt}{3ex}\minrev{OpenSSL}} \\ \hline
        \rule{0pt}{3ex}\minrev{\ourbilstm} & \minrev{0.37 (0.31)} & \minrev{0.68 (0.68)} & \minrev{0.75 (0.74)} & \minrev{0.62 (0.61)} \\ \hline
        \rule{0pt}{3ex}\minrev{\our} & \bf{\minrev{0.41 (0.31)}} & \bf{\minrev{0.73 (0.68)}} & \bf{\minrev{0.81 (0.78)}} & \bf{\minrev{0.67 (0.62)}} \\ \hline
        \end{tabular}
  \end{center}
\end{table}

Tables~\ref{tab:discussion_clean_results} and~\ref{tab:discussion_realistic_results} show the average and median performance statistics of \ourbilstm in \cleanset and \realset, respectively. We also mention the results of \our for comparison. On average, in \cleanset, \ourbilstm achieved 0.73 MCC, 0.84 F-1, 0.84 Precision, and 0.84 Recall for Linux Kernel releases; 0.54 MCC, 0.72 F-1, 0.85 Precision, and 0.63 Recall for Wireshark releases; and 0.71 MCC, 0.82 F-1, 0.95 Precision, and 0.73 Recall for OpenSSL releases. In \realset, \ourbilstm achieved 0.38 MCC, 0.65 F-1, 0.84 Precision, and 0.53 Recall for Linux Kernel releases; 0.34 MCC, 0.66 F-1, 0.73 Precision, and 0.61 Recall for Wireshark releases; and 0.37 MCC, 0.68 F-1, 0.75 Precision, and 0.62 Recall for OpenSSL releases.

\begin{figure*}[t]
  \begin{center}
    \makebox[\textwidth]{\includegraphics[width=\textwidth]{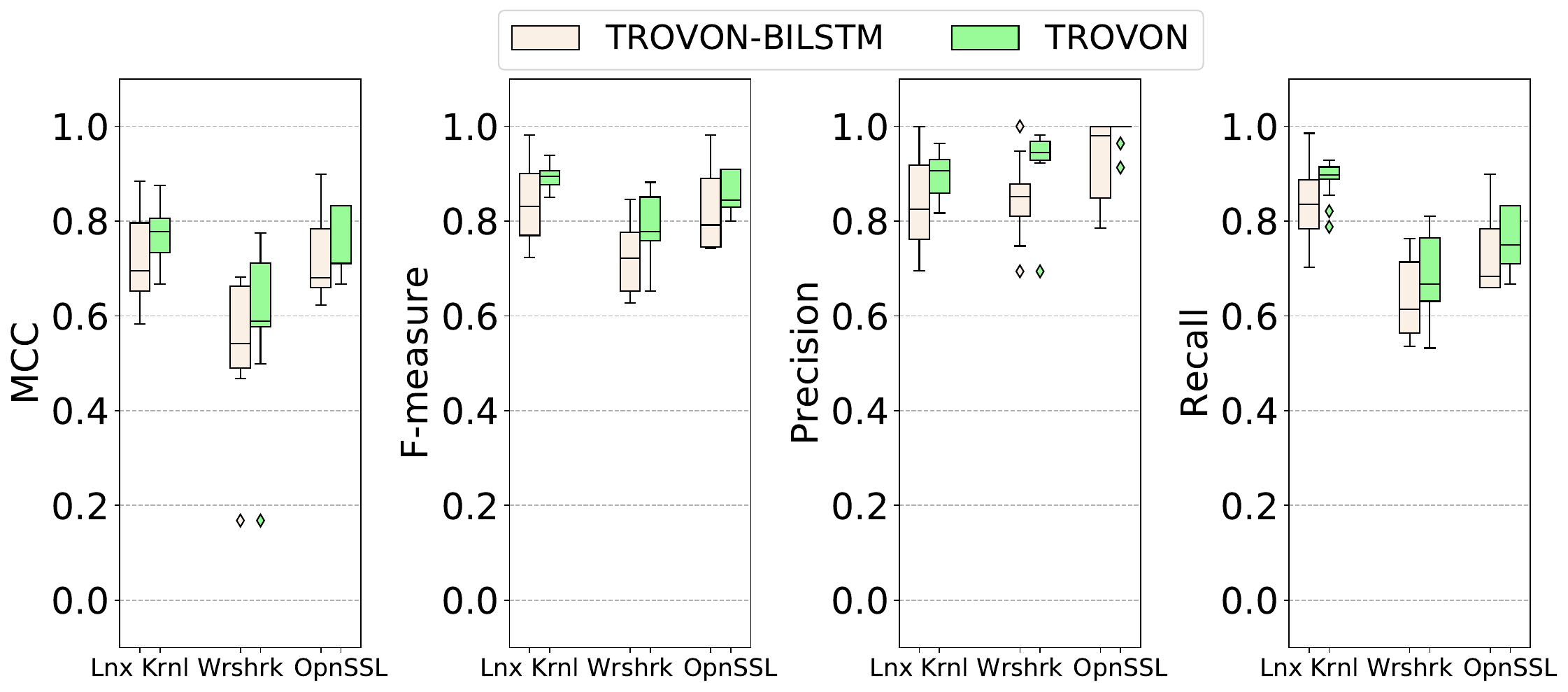}}
    \caption{\minrev{Comparison between \ourbilstm and \our under \cleanset: \our outperforms \ourbilstm by 6.49\% in MCC, 6.63\% in F-1, 6.40\% in Precision, and 6.75\% in Recall.}
    }
    \label{fig:discussion_clean_results}
  \end{center}
\vspace{-1em}
\end{figure*}

\begin{figure*}[t]
  \begin{center}
    \makebox[\textwidth]{\includegraphics[width=\textwidth]{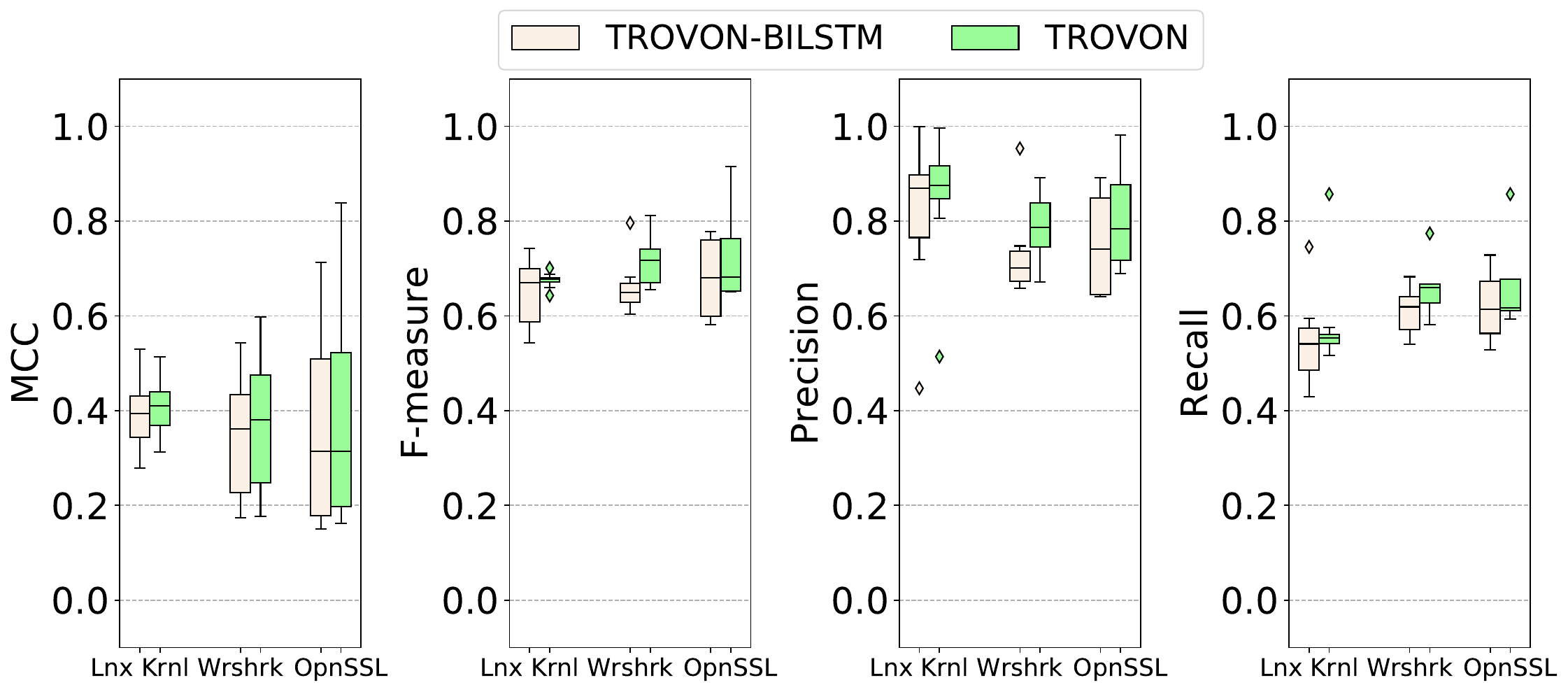}}
    \caption{\minrev{Comparison between \ourbilstm and \our under \realset: \our outperforms \ourbilstm by 5.08\% in MCC, 5.21\% in F-1, 5.01\% in Precision, and 5.43\% in Recall.}
    }
    \label{fig:discussion_realistic_results}
  \end{center}
\vspace{-1em}
\end{figure*}

Figures~\ref{fig:discussion_clean_results} and~\ref{fig:discussion_realistic_results} show the performance comparison of \ourbilstm and \our in \cleanset and \realset, respectively. The figures show that \our performs comparatively better than \ourbilstm. Overall, when trained with vulnerabilities, in \cleanset, \our outperforms \ourbilstm by 6.49\% in MCC, 6.63\% in F-1, 6.40\% in Precision, and 6.75\% in Recall. In \realset, \our outperforms \ourbilstm by 5.08\% in MCC, 5.21\% in F-1, 5.01\% in Precision, and 5.43\% in Recall.
}

\section{Threats To Validity}
\label{sec:threats-to-validity}

\textit{Construct Validity}: 
\norev{We use \emph{VulData7}~\cite{JimenezPT18} for data collection using the Git commit IDs provided in the CVE-NVD database. 
This process ensures the retrieval of known and fixed vulnerabilities, whereas undiscovered or unfixed vulnerabilities are ignored. 
This may result in false negatives with a potential impact on our measurements. 
However, given the size of Linux Kernel, Wireshark, and OpenSSL and their long history of vulnerability reports, we believe that it is unlikely to have many such cases.}

Another concern originates from our choice to learn from the vulnerable and fixed pairs of components. 
Since \our has access to this information one can argue that the improved performance is due to this additional knowledge of fixed components.
To diminish this concern we also included the fixed versions of the vulnerable files in the training set for training existing techniques, but this resulted in negligible differences in their performance. 

One may wonder if most of the vulnerabilities are introduced due to code changes performed between the releases and whether every changed component between adjacent releases can be flagged as vulnerable. 
We analyzed our data and found that the results are close to random guessing \rev{with} MCC- \rev{0.06, 0.09, 0.1} and Precision- \rev{0.04, 0.08, 0.14} for Linux Kernel, Wireshark, and OpenSSL project releases, \rev{respectively}. 
These results are in accordance with the findings of Jimenez et al.~\cite{Jimenez+2019} that most vulnerabilities span across multiple releases without being detected, and mislead the predictions, \eg an existing vulnerability in release \emph{\rev{R1}} may get detected and fixed in the release \emph{{\rev{R4}}}. 
Also, \rev{many files} are modified between the releases, \ie 29.95\%, 72.53\%, and 73.58\% \rev{of the files, on average, are changed for} Linux Kernel, Wireshark, and OpenSSL, \rev{which} adds to the imprecision of this baseline by producing excessive numbers of false positives/negatives. 

\textit{Internal Validity}: 
\norev{We do not consider non-vulnerable components for training as these files can in fact be vulnerable (vulnerability undetected till date) and may mislead our predictor.
Still, we train on the unchanged and fixed parts of the vulnerable components as we believe that these are unlikely to be vulnerable. To support this intuition, we checked our data and found that it is indeed true, \ie components having more than one vulnerability, with one fixed and the other not, are on average 0.037\%, 0.19\%, and 0.24\% of the Linux Kernel, Wireshark, and OpenSSL components per release.}

\rev{We use \emph{FrameVPM}~\cite{Jimenez+2019} to implement vulnerability prediction models for \sm, \im, \fc, and \tm. As none of the replicated approaches provide a replication package, the framework may not have implemented precisely the original approaches. To reduce this threat we inspected the code, parameters, and experiment decisions to perform the most accurate replication possible. Given that our results are in line with the previous replication studies~\cite{jimenez2016empirical,Jimenez+2019} and the original studies~\cite{ShinMWO2011,Neuhaus+2007}, we believe this threat is of less significance.

Similarly, we implement \dev, \lstm, and \lstmrf based on our understanding of authors' work described in the available articles because the author's implementation/source-code of these approaches is not available. Still, there is a possibility that we may not have implemented the original approaches as precisely as the authors of these approaches would have. Nevertheless, these approaches make the clean labeling assumption \cite{Jimenez+2019} thereby experimenting fundamental limitations on their performance. This is actually the key reason why previous work reports much better results. Nevertheless, when using \cleanset, we found F-1 scores of 32.73\% and 36.54\% for Linux Kernel and Wireshark, which are in line with the results reported by Zhou \etal~\cite{zhou2019devign} (\ie F-1 score of 24.64\% and 42.05\% for Linux Kernel and Wireshark), in their case of imbalanced data (the only case that is somehow comparable with our analysis).}

\textit{External Validity}: Although the study expands its evaluation to three security-critical open source systems, the results may not generalize to other projects (e.g., Android). 
Additional studies are required to sufficiently take care of the generalization threat.
Also, we split the methods into sequences of no more than \textit{50} tokens each.
Method-splitting in larger sequences may require more training time and computational resources but can lead to better results.

\section{Related Work}
\label{sec:related-work}
Early work in the area of vulnerability prediction has focused on defining features that could be linked to vulnerabilities and thus to be used to train learners.
The first such work can be traced back to the study of Neuhaus \etal~\cite{Neuhaus+2007}, which investigated the use of libraries and function calls. Later, Shin \etal~\cite{shin_can_2013,ShinMWO2011} and Zulkernine \etal~\cite{10.1016/j.sysarc.2010.06.003} investigated the use of code metrics such as complexity, code churn, and object oriented metrics. Theisen and Williams \cite{TheisenW20} showed that a combination of these features can slightly improve the F-score and recommend identifying new features. 

These approaches, although promising, were all using features designed based on human intuition.  Scandariato \etal~\cite{6860243} advocated that the learners should find their features without human intervention. To achieve this, they suggested the \tm approach where code is treated as text and the learner learns from \emph{Bag of Words} (BoW). The results of their exploratory study demonstrated that \tm's prediction power was superior to the state of the art vulnerability prediction models with good performance for both precision and recall in intra-project predictions.

Recently, deep learning techniques have been explored to automatically learn the required features to predict vulnerabilities. 
Li et. al \cite{DBLP:conf/ndss/LiZXO0WDZ18} used Bidirectional LSTMs to train a vulnerability prediction model on \emph{code gadgets}, which are semantically related lines of code.
Under \rev{\cleanset}, this technique was shown to be effective for analyzing two particular weaknesses, namely, buffer error vulnerabilities (CWE-119) and management error vulnerabilities (CWE-399).
In contrast, \our trains the translation model on sequences extracted from the source code and does not target specific weaknesses.

Machine learning has also been used in other software engineering prediction tasks.
For instance, several works~\cite{DAmbros+2012,Hall+2012,Yang+2015,Wang+2016} used machine learning models for defect prediction.
Particularly, RNN models have been used for automatically fixing errors in C programs~\cite{Gupta+2017},
for generating API usage sequences~\cite{Gu+2016}, and for fault localization~\cite{Huo+2016}.
Closer to our work, machine translation-based approaches have been successfully applied to automatically learn code features for detecting code clones~\cite{White+2016}, \rev{and interesting mutants~\cite{garg2022cerebro},} 
for learning how to mutate source code from bugs~\cite{Tufano_2019}, and \rev{to produce} bug-fixing repairs~\cite{DBLP:journals/tosem/TufanoWBPWP19}.
To our knowledge, \our is the first approach that proposes and evaluates a machine translation-based vulnerability prediction.

\section{Conclusion}
\label{sec:conclusion}
This paper proposes \rev{\our}, a machine translation based approach to automatically learn to predict vulnerable components from noisy historical data. \rev{Taking advantage of the large amounts of historical data, our} predictions can be used to assist developers in code reviews and security testing. The important advantage of \our is that it is completely automatic as it learns latent features (context, patterns, etc.) linked with vulnerabilities based on information mining from code repositories (in particular by analyzing historical vulnerability fixes and their context). 
We empirically evaluated the effectiveness of \our following the methodological guidelines set by Jimenez et al. \cite{Jimenez+2019}. 
In particular, we demonstrated that \our can mitigate the problem of real-world noisy data on the releases of the three security-critical open source systems that were used by previous research. 
Moreover, we showed that \our outperforms existing techniques under both, clean and realistic, (\ie noisy) training data \rev{settings}. 
On average, when trained on clean data, \our achieved an overall improvement of \rev{80.39\%} in MCC score.
\rev{Moreover, in \realset, \our achieved 3.63 times higher MCC score in comparison to existing approaches.}


	\section*{Conflict of interest}
	
	The authors declare that they have no conflict of interest.
%
%
%

\bibliographystyle{plain}
\bibliography{Bibliography}

\begin{thebibliography}{10}

\bibitem{CVE:Terminology}
Definition of vulnerability.
\newblock \url{https://cve.mitre.org/about/terminology.html}, (accessed May 01,
  2021).

\bibitem{heartbleed}
The heartbleed bug.
\newblock \url{https://heartbleed.com/}, (accessed May 01, 2021).

\bibitem{linux-news}
Linux in 2020: 27.8 million lines of code in the kernel.
\newblock
  \url{https://www.linux.com/news/linux-in-2020-27-8-million-lines-of-code-in-the-kernel-1-3-million-in-systemd/},
  (accessed May 01, 2021).

\bibitem{LinuxKernel}
Linux kernal.
\newblock \url{https://www.kernel.org}, (accessed May 01, 2021).

\bibitem{NVD}
National vulnerability database.
\newblock \url{https://nvd.nist.gov}, (accessed May 01, 2021).

\bibitem{OpenSSL}
Openssl.
\newblock \url{https://www.openssl.org}, (accessed May 01, 2021).

\bibitem{Vulnerabilities}
Vulnerabilities.
\newblock \url{https://owasp.org/www-community/vulnerabilities/}, (accessed May
  01, 2021).

\bibitem{WireShark}
Wireshark.
\newblock \url{https://www.wireshark.org}, (accessed May 01, 2021).

\bibitem{bahdanau2014neural}
Dzmitry Bahdanau, Kyunghyun Cho, and Yoshua Bengio.
\newblock Neural machine translation by jointly learning to align and
  translate, 2014.

\bibitem{Britz:2017}
D.~{Britz}, A.~{Goldie}, T.~{Luong}, and Q.~{Le}.
\newblock {Massive Exploration of Neural Machine Translation Architectures}.
\newblock {\em ArXiv e-prints}, March 2017.

\bibitem{rnnencoderdecoder}
Jason Brownlee.
\newblock Encoder-decoder recurrent neural network models for neural machine
  translation.
\newblock
  \url{https://machinelearningmastery.com/encoder-decoder-recurrent-neural-network-models-neural-machine-translation/},
  2018 (accessed February 01, 2022).

\bibitem{lstmcomparisonwebsite}
Jason Brownlee.
\newblock When to use mlp, cnn, and rnn neural networks.
\newblock
  \url{https://machinelearningmastery.com/when-to-use-mlp-cnn-and-rnn-neural-networks},
  2018 (accessed May 01, 2021).

\bibitem{10.1016/j.sysarc.2010.06.003}
Istehad Chowdhury and Mohammad Zulkernine.
\newblock Using complexity, coupling, and cohesion metrics as early indicators
  of vulnerabilities.
\newblock {\em J. Syst. Archit.}, 57(3):294–313, March 2011.

\bibitem{7816536}
M.~L. {Collard} and J.~I. {Maletic}.
\newblock srcml 1.0: Explore, analyze, and manipulate source code.
\newblock In {\em 2016 IEEE International Conference on Software Maintenance
  and Evolution (ICSME)}, pages 649--649, 2016.

\bibitem{Dam+2018}
H.~K. {Dam}, T.~{Tran}, T.~T.~M. {Pham}, S.~W. {Ng}, J.~{Grundy}, and
  A.~{Ghose}.
\newblock Automatic feature learning for predicting vulnerable software
  components.
\newblock {\em IEEE Transactions on Software Engineering}, pages 1--1, 2018.

\bibitem{DAmbros+2012}
Marco D'Ambros, Michele Lanza, and Romain Robbes.
\newblock Evaluating defect prediction approaches: A benchmark and an extensive
  comparison.
\newblock {\em Empirical Softw. Engg.}, 17(4–5):531–577, August 2012.

\bibitem{tensorflow2015-whitepaper}
Mart\'{\i}n~Abadi et~al.
\newblock {TensorFlow}: Large-scale machine learning on heterogeneous systems,
  2015.
\newblock Software available from tensorflow.org.

\bibitem{falleri:hal-01054552}
Jean-R{\'e}my Falleri, Flor{\'e}al Morandat, Xavier Blanc, Matias Martinez, and
  Martin Monperrus.
\newblock {Fine-grained and Accurate Source Code Differencing}.
\newblock In {\em {Proceedings of the International Conference on Automated
  Software Engineering}}, pages 313--324, V{\"a}steras, Sweden, 2014.
\newblock update for oadoi on Nov 02 2018.

\bibitem{garg2022cerebro}
Aayush Garg, Milos Ojdanic, Renzo Degiovanni, Thierry~Titcheu Chekam, Mike
  Papadakis, and Yves Le~Traon.
\newblock Cerebro: Static subsuming mutant selection.
\newblock {\em IEEE Transactions on Software Engineering}, pages 1--1, 2022.

\bibitem{Gu+2016}
Xiaodong Gu, Hongyu Zhang, Dongmei Zhang, and Sunghun Kim.
\newblock Deep api learning.
\newblock In {\em Proceedings of the 2016 24th ACM SIGSOFT International
  Symposium on Foundations of Software Engineering}, FSE 2016, page 631–642,
  New York, NY, USA, 2016. Association for Computing Machinery.

\bibitem{Gupta+2017}
Rahul Gupta, Soham Pal, Aditya Kanade, and Shirish Shevade.
\newblock Deepfix: Fixing common c language errors by deep learning.
\newblock In {\em Proceedings of the Thirty-First AAAI Conference on Artificial
  Intelligence}, AAAI’17, page 1345–1351. AAAI Press, 2017.

\bibitem{Hall+2012}
T.~{Hall}, S.~{Beecham}, D.~{Bowes}, D.~{Gray}, and S.~{Counsell}.
\newblock A systematic literature review on fault prediction performance in
  software engineering.
\newblock {\em IEEE Transactions on Software Engineering}, 38(6):1276--1304,
  2012.

\bibitem{HochreiterHochreiter1997}
Sepp Hochreiter and J\"{u}rgen Schmidhuber.
\newblock Long short-term memory.
\newblock {\em Neural Comput.}, 9(8):1735–1780, November 1997.

\bibitem{Huo+2016}
Xuan Huo, Ming Li, and Zhi-Hua Zhou.
\newblock Learning unified features from natural and programming languages for
  locating buggy source code.
\newblock In {\em Proceedings of the Twenty-Fifth International Joint
  Conference on Artificial Intelligence}, IJCAI’16, page 1606–1612. AAAI
  Press, 2016.

\bibitem{jimenez2016empirical}
Matthieu Jimenez, Mike Papadakis, and Yves Le~Traon.
\newblock An empirical analysis of vulnerabilities in openssl and the linux
  kernel.
\newblock In {\em 2016 23rd Asia-Pacific Software Engineering Conference
  (APSEC)}, pages 105--112. IEEE, 2016.

\bibitem{JimenezPT18}
Matthieu Jimenez, Mike Papadakis, and Yves~Le Traon.
\newblock Enabling the continous analysis of security vulnerabilities with
  vuldata7.
\newblock In {\em Proceedings of the 18th IEEE International Working Conference
  on Source Code Analysis and Manipulation {SCAM} 2018, Madrid, Spain,
  September 23-24, 2018}, 2018.

\bibitem{Jimenez+2019}
Matthieu Jimenez, Renaud Rwemalika, Mike Papadakis, Federica Sarro, Yves
  Le~Traon, and Mark Harman.
\newblock The importance of accounting for real-world labelling when predicting
  software vulnerabilities.
\newblock In {\em Proceedings of the 2019 27th ACM Joint Meeting on European
  Software Engineering Conference and Symposium on the Foundations of Software
  Engineering}, ESEC/FSE 2019, page 695–705, New York, NY, USA, 2019.
  Association for Computing Machinery.

\bibitem{10.5555/1643031.1643034}
Igor Kononenko.
\newblock On biases in estimating multi-valued attributes.
\newblock In {\em Proceedings of the 14th International Joint Conference on
  Artificial Intelligence - Volume 2}, IJCAI’95, page 1034–1040, San
  Francisco, CA, USA, 1995. Morgan Kaufmann Publishers Inc.

\bibitem{DBLP:conf/ndss/LiZXO0WDZ18}
Zhen Li, Deqing Zou, Shouhuai Xu, Xinyu Ou, Hai Jin, Sujuan Wang, Zhijun Deng,
  and Yuyi Zhong.
\newblock Vuldeepecker: {A} deep learning-based system for vulnerability
  detection.
\newblock In {\em 25th Annual Network and Distributed System Security
  Symposium, {NDSS} 2018, San Diego, California, USA, February 18-21, 2018},
  2018.

\bibitem{MATTHEWS1975442}
B.W. Matthews.
\newblock Comparison of the predicted and observed secondary structure of t4
  phage lysozyme.
\newblock {\em Biochimica et Biophysica Acta (BBA) - Protein Structure},
  405(2):442 -- 451, 1975.

\bibitem{Morrison+2015}
Patrick Morrison, Kim Herzig, Brendan Murphy, and Laurie Williams.
\newblock Challenges with applying vulnerability prediction models.
\newblock In {\em Proceedings of the 2015 Symposium and Bootcamp on the Science
  of Security}, HotSoS ’15, New York, NY, USA, 2015. Association for
  Computing Machinery.

\bibitem{DBLP:conf/sac/MoshtariS16}
Sara Moshtari and Ashkan Sami.
\newblock Evaluating and comparing complexity, coupling and a new proposed set
  of coupling metrics in cross-project vulnerability prediction.
\newblock In Sascha Ossowski, editor, {\em Proceedings of the 31st Annual {ACM}
  Symposium on Applied Computing, Pisa, Italy, April 4-8, 2016}, pages
  1415--1421. {ACM}, 2016.

\bibitem{Neuhaus+2007}
Stephan Neuhaus, Thomas Zimmermann, Christian Holler, and Andreas Zeller.
\newblock Predicting vulnerable software components.
\newblock In {\em Proceedings of the 14th ACM Conference on Computer and
  Communications Security}, CCS ’07, page 529–540, New York, NY, USA, 2007.
  Association for Computing Machinery.

\bibitem{PotterMcGraw2004}
B.~{Potter} and G.~{McGraw}.
\newblock Software security testing.
\newblock {\em IEEE Security Privacy}, 2(5):81--85, 2004.

\bibitem{6860243}
R.~{Scandariato}, J.~{Walden}, A.~{Hovsepyan}, and W.~{Joosen}.
\newblock Predicting vulnerable software components via text mining.
\newblock {\em IEEE Transactions on Software Engineering}, 40(10):993--1006,
  2014.

\bibitem{6824804}
M.~{Shepperd}, D.~{Bowes}, and T.~{Hall}.
\newblock Researcher bias: The use of machine learning in software defect
  prediction.
\newblock {\em IEEE Transactions on Software Engineering}, 40(6):603--616,
  2014.

\bibitem{lstmcomparison2019}
Apeksha Shewalkar, Deepika Nyavanandi, and Simone Ludwig.
\newblock Performance evaluation of deep neural networks applied to speech
  recognition: Rnn, lstm and gru.
\newblock {\em Journal of Artificial Intelligence and Soft Computing Research},
  9:235--245, 10 2019.

\bibitem{ShinMWO2011}
Yonghee Shin, Andrew Meneely, Laurie Williams, and Jason~A. Osborne.
\newblock Evaluating complexity, code churn, and developer activity metrics as
  indicators of software vulnerabilities.
\newblock {\em IEEE Trans. Softw. Eng.}, 37(6):772–787, November 2011.

\bibitem{10.1145/1414004.1414065}
Yonghee Shin and Laurie Williams.
\newblock An empirical model to predict security vulnerabilities using code
  complexity metrics.
\newblock In {\em Proceedings of the Second ACM-IEEE International Symposium on
  Empirical Software Engineering and Measurement}, ESEM ’08, page 315–317,
  New York, NY, USA, 2008. Association for Computing Machinery.

\bibitem{shin_can_2013}
Yonghee Shin and Laurie Williams.
\newblock Can traditional fault prediction models be used for vulnerability
  prediction?
\newblock {\em Empirical Software Engineering}, 18(1):25--59, February 2013.

\bibitem{sutskever2014sequence}
Ilya Sutskever, Oriol Vinyals, and Quoc~V. Le.
\newblock Sequence to sequence learning with neural networks, 2014.

\bibitem{TangZYLZX15}
Yaming Tang, Fei Zhao, Yibiao Yang, Hongmin Lu, Yuming Zhou, and Baowen Xu.
\newblock Predicting vulnerable components via text mining or software metrics?
  an effort-aware perspective.
\newblock In {\em QRS}, pages 27--36. IEEE, 2015.

\bibitem{TheisenW20}
Christopher Theisen and Laurie~A. Williams.
\newblock Better together: Comparing vulnerability prediction models.
\newblock {\em Inf. Softw. Technol.}, 119, 2020.

\bibitem{Tufano_2019}
Michele Tufano, Cody Watson, Gabriele Bavota, Massimiliano Di~Penta, Martin
  White, and Denys Poshyvanyk.
\newblock Learning how to mutate source code from bug-fixes.
\newblock {\em 2019 IEEE International Conference on Software Maintenance and
  Evolution (ICSME)}, Sep 2019.

\bibitem{DBLP:journals/tosem/TufanoWBPWP19}
Michele Tufano, Cody Watson, Gabriele Bavota, Massimiliano~Di Penta, Martin
  White, and Denys Poshyvanyk.
\newblock An empirical study on learning bug-fixing patches in the wild via
  neural machine translation.
\newblock {\em {ACM} Trans. Softw. Eng. Methodol.}, 28(4):19:1--19:29, 2019.

\bibitem{VarghaDelaney2000}
András Vargha and Harold~D. Delaney.
\newblock A critique and improvement of the "cl" common language effect size
  statistics of mcgraw and wong.
\newblock {\em Journal of Educational and Behavioral Statistics},
  25(2):101--132, 2000.

\bibitem{Wang+2016}
Song Wang, Taiyue Liu, and Lin Tan.
\newblock Automatically learning semantic features for defect prediction.
\newblock In {\em Proceedings of the 38th International Conference on Software
  Engineering}, ICSE ’16, page 297–308, New York, NY, USA, 2016.
  Association for Computing Machinery.

\bibitem{White+2016}
M.~{White}, M.~{Tufano}, C.~{Vendome}, and D.~{Poshyvanyk}.
\newblock Deep learning code fragments for code clone detection.
\newblock In {\em 2016 31st IEEE/ACM International Conference on Automated
  Software Engineering (ASE)}, pages 87--98, 2016.

\bibitem{Wilcoxon1945}
Frank Wilcoxon.
\newblock Individual comparisons by ranking methods.
\newblock {\em Biometrics Bulletin}, 1(6):80--83, 1945.

\bibitem{Yang+2015}
X.~{Yang}, D.~{Lo}, X.~{Xia}, Y.~{Zhang}, and J.~{Sun}.
\newblock Deep learning for just-in-time defect prediction.
\newblock In {\em 2015 IEEE International Conference on Software Quality,
  Reliability and Security}, pages 17--26, 2015.

\bibitem{zhou2019devign}
Yaqin Zhou, Shangqing Liu, Jingkai Siow, Xiaoning Du, and Yang Liu.
\newblock Devign: Effective vulnerability identification by learning
  comprehensive program semantics via graph neural networks, 2019.

\bibitem{Zimmermann+2009}
Thomas Zimmermann, Nachiappan Nagappan, Harald Gall, Emanuel Giger, and Brendan
  Murphy.
\newblock Cross-project defect prediction: A large scale experiment on data vs.
  domain vs. process.
\newblock In {\em Proceedings of the 7th Joint Meeting of the European Software
  Engineering Conference and the ACM SIGSOFT Symposium on The Foundations of
  Software Engineering}, ESEC/FSE ’09, page 91–100, New York, NY, USA,
  2009. Association for Computing Machinery.

\end{thebibliography}

\end{document}